\documentclass[sigconf]{acmart} %% CCS: DO NOT REMOVE
\renewcommand\footnotetextcopyrightpermission[1]{}

\AtBeginDocument{%
  }

\acmConference[Preprint]{}{}{}

\usepackage{tikz}
\usepackage{url}
\usepackage{amsmath}
\usepackage{xspace}
\usepackage[ruled,vlined,linesnumbered]{algorithm2e}
\usepackage{hyperref}
\usepackage{subcaption}
\usepackage{graphicx}
\usepackage{etoolbox}
\usepackage{enumitem}
\usepackage{multirow}
\usepackage{seqsplit}
\usepackage{xcolor}
\usepackage{tcolorbox}
\usepackage{mathtools}
\usepackage{booktabs}
\usepackage{circledsteps}
\usepackage{multicol}
\usepackage{minted}

\makeatletter
\renewcommand{\nllabel}[1]
 {{\let\@currentlabel\algocf@currentlabel
  \let\@currentcounter\algocf@currentcounter
  \label{#1}}}%

\renewcommand{\algocf@nl@sethref}[1]{%
  \renewcommand{\theHAlgoLine}{\thealgocfproc.#1}%
  \hyper@refstepcounter{AlgoLine}%
  \gdef\algocf@currentlabel{#1}%
  \gdef\algocf@currentcounter{AlgoLine}%
 }%
\makeatother

\usepackage[capitalise,noabbrev]{cleveref}

\newcommand{\kw}[1]{\ensuremath{\mathtt{#1}}}

\newcommand{\tname}{\ensuremath{t}}
\newcommand{\pname}{\ensuremath{p}}
\newcommand{\vtype}{\ensuremath{s}}
\newcommand{\vvalue}{\ensuremath{v}}

\definecolor{Cyan4}{RGB}{0,139,139}
\definecolor{Orange4}{RGB}{139,90,0}

\newsavebox{\largestimage}
\makeatletter
\newcommand{\algorithmfootnote}[2][\footnotesize]{%
  \let\old@algocf@finish\@algocf@finish% Store algorithm finish macro
  \def\@algocf@finish{\old@algocf@finish% Update finish macro to insert "footnote"
    \leavevmode\rlap{\begin{minipage}{\linewidth}
    #1#2
    \end{minipage}}%
  }%
}
\makeatother

\makeatletter
\patchcmd{\algocf@makecaption@ruled}{\hsize}{\columnwidth}{}{} % Caption to stretch full text width
\patchcmd{\@algocf@start}{-1.5em}{0em}{}{} % For // to right margin
\makeatother

\newtcolorbox{mybox}[0]
{
  colframe = black,
}

\newcommand{\system}{Progent\xspace}

\newcommand{\agent}{\ensuremath{\mathcal{A}}{}}
\newcommand{\user}{\ensuremath{\mathcal{U}}{}}
\newcommand{\env}{\ensuremath{\mathcal{E}}{}}
\newcommand{\tools}{\ensuremath{\mathcal{T}}{}}
\newcommand{\tool}{\ensuremath{T}{}}
\newcommand{\obs}{\ensuremath{o}{}}

\newcommand{\call}{\ensuremath{c}{}}

\newcommand{\policy}{\ensuremath{P}{}}

\newcommand{\effect}{\ensuremath{E}{}}
\newcommand{\expr}{\ensuremath{e}{}}

\newcommand{\fallback}{\ensuremath{f}{}}
\newcommand{\bop}{\ensuremath{\mathit{bop}}{}}
\newcommand{\msg}{\ensuremath{\mathit{msg}}{}}

\newcommand{\secgpt}{IsolateGPT}
\newcommand{\iflow}{f-secure}

\newcommand{\myhl}[1]{\textcolor{Cyan4}{#1}}

\begin{document}

%% allowing the author to define a "short title" to be used in page
%% headers.

\title{\system: Securing AI Agents with Privilege Control} %% CCS: you MUST provide a title


%% CCS: at submission time, the submission MUST be anonymized. Hence
%% authors MUST be commented out.

\author{
Tianneng Shi\textsuperscript{1}, Jingxuan He\textsuperscript{1}, Zhun Wang\textsuperscript{1}, Hongwei Li\textsuperscript{2}, Linyu Wu\textsuperscript{3}, Wenbo Guo\textsuperscript{2}, Dawn Song\textsuperscript{1}  \\
\textsuperscript{1}UC Berkeley \textsuperscript{2}UC Santa Barbara \textsuperscript{3}National University of Singapore
}

\renewcommand{\shortauthors}{Trovato et al.}

\begin{abstract}
AI agents interact with external environments through tool calls, exposing them to attacks like indirect prompt injection that can trigger unauthorized actions.
Securing these agents is challenging: they behave autonomously and probabilistically, security requirements evolve depending on the user's task and execution state, and there is an inherent tradeofff between security and utility.

In this work, we introduce \system{}, a novel framework that secures AI agents via privilege control.
\system{} represents privilege as a security policy consisting of symbolic rules over tool names and arguments.
These rules specify which tool calls are allowed for task completion and which unnecessary ones are blocked for security.
Every tool call is checked against such a policy through a deterministic procedure, enforcing the principle of least privilege.
To handle diverse user tasks and evolving execution contexts, an LLM automatically generates the initial policy from the user's task and updates it during execution as new information arrives.
Each proposed update is determined by an SMT solver to be either a narrowing (applied automatically) or an expansion (requiring explicit approval), ensuring that the agent's effective action space can only shrink without approval (monotonic confinement).
This deterministic update mechanism preserves utility and prevents silent privilege escalation, even when adversarial inputs are present.

Our evaluation on popular benchmarks (i.e., AgentDojo and ASB) shows that \system{} significantly reduces attack success rates while maintaining high utility.
We further validate \system{}'s practicality by showcasing its effectiveness in real-world agent frameworks such as LangChain and OpenAI Agents SDK.
\end{abstract}


% \keywords{AI Agent, Security, Prompt Injection, Privilege Control} %% CCS: DO NOT REMOVE but you MAY update


\maketitle

% \section{Introduction} %% CCS: You MAY change the title and,
                       %% obviously, add text, sections, figures,
                       %% tables, etc. 
%% CCS: For help and more latex examples, refer to
%% `sample-sigconf.tex', provided in the distribution
%% https://portalparts.acm.org/hippo/latex_templates/acmart-primary.zip 
%%


%% CCS: to preserve anonymity, NO acknowledgements to fundings, projects or persons should be used at
%% submission time
%% CCS: this section MAY be used to acknowledge the use of AI when used only for minor editorial improvements (e.g., grammar, spelling, or light style polishing) 
% \begin{acks}
% This paper was edited for grammar using [Tool Name].
% \end{acks}

\section{Introduction}
\label{sec:intro}

AI agents have emerged as a promising platform for general and autonomous task solving~\cite{wang2024survey,shi2024ehragent,wang2024executable,yao2022react}.
At the core of these agents is a large language model (LLM), which interacts with the external environment through diverse sets of tools~\cite{schick2023toolformer,qin2023toolllm}.
For instance, a personal assistant agent uses email toolkits~\cite{gmail-langchain} to manage emails, and a coding agent uses code interpreters and the command line~\cite{wang2024executable}.

\paragraph{Security Risks in AI Agents}
The ability to interact with external environments makes AI agents powerful, enabling them to perform diverse tasks across domains such as communication, scheduling, and finance.
However, this same ability also creates a large attack surface.
Recent research has raised serious security concerns about AI agents~\cite{li2024personal,he2024emerged,wu2024new}, where attackers exploit the agent's interaction with the environment by injecting adversarial instructions into external data sources.
When the agent retrieves such data via tool calls, the injected instructions can redirect the agent to perform dangerous actions, such as unauthorized financial transactions~\cite{debenedetti2024agentdojo} or data exfiltration~\cite{liao2024eia}.
This class of attacks is known as \emph{indirect prompt injection}~\cite{greshake2023not,liu2023prompt}.

\paragraph{Challenges for Securing AI Agents}
Securing AI agents is challenging for several reasons.
First, agent behavior is inherently non-deterministic: an LLM processes unstructured natural language and may make different decisions across runs.
In contrast, security requires deterministic enforcement that holds regardless of how the LLM reasons, even under adversarial influence.
Reconciling these two worlds is non-trivial.
Second, agent security depends on both the user's task and the information the agent gathers during execution.
Different tasks require different tools and arguments, and the relevant context evolves at runtime as the agent interacts with the environment.
A defense must track these changes automatically, since involving the user in every security decision is impractical and negates the benefit of an autonomous agent.
Finally, agents must sometimes expand their privileges to complete benign tasks.
Yet, adversaries also seek to exploit the same mechanism to induce harmful actions, and both legitimate and adversarial expansion requests arrive through the same channel.
Distinguishing legitimate from adversarial expansions, without sacrificing security or utility, is a fundamental challenge.
We elaborate on these challenges with a running example in \cref{sec:overview}.

\paragraph{Our Work: \system{}}
We introduce \system{}, a novel framework that secures AI agents via privilege control.
Our key idea is to enforce the principle of least privilege: allow only tool calls essential for task completion and block unnecessary ones.
We focus on tool calls because all external effects of an agent occur through them, and they expose a structured interface (tool names and typed arguments), unlike the agent's unstructured reasoning.
This makes tool calls a natural point for deterministic security enforcement.
\system{} leverages a privilege control policy consisting of symbolic rules over tool names and arguments.
These rules specify which tool calls are allowed and which should be blocked.
Every tool call is checked against such a policy, which deterministically decides whether it is permitted.

To handle context-dependent security requirements,  \system{} utilizes an LLM to automatically generate the initial policy from the user's task and updates it as new information arrives during execution.
This allows security enforcement to adapt to both task and execution context while minimizing manual effort.
However, introducing an LLM for policy construction introduces non-determinism, creating a fundamental tension between automation and deterministic enforcement.

To address this, we further propose that each policy update be checked by a deterministic SMT solver as either a narrowing (applied automatically) or an expansion (requiring explicit approval).
This ensures that the agent's effective action space can only narrow without approval, a property we call monotonic confinement.
As a result, even if the LLM is manipulated by adversarial inputs, this deterministic check prevents any silent privilege escalation.
At the same time, it preserves utility by allowing controlled privilege expansion when explicitly authorized.

\paragraph{Implementation and Evaluation}
Since \system{} operates at the tool-call level, it can be integrated by inserting a separate module between the agent and its tools.
This requires no changes to the agent's internal architecture, making integration non-intrusive and generalizable across diverse agent frameworks.

We evaluate \system{} on AgentDojo~\cite{debenedetti2024agentdojo} and ASB~\cite{zhang2024agent}.
\system{} reduces the attack success rate (ASR) from 39.9\% to 1.0\% on AgentDojo, and from 70.3\% to 3.9\% on ASB, while maintaining the agent's utility in both cases.
We further conduct extensive ablation studies to analyze key components of \system{}, including model choices and policy update approval configurations.
Finally, we extend AgentDojo into an MCP-based benchmark and connect it to various real-world agent frameworks (e.g., LangChain and OpenAI Agents SDK), where we demonstrate the effectiveness of \system{} on protecting both single- and multi-agent systems.

\paragraph{Main Contributions}
In summary, our main contributions are:
\begin{itemize}[leftmargin=*, itemsep=0pt, parsep=1pt, topsep=1pt]
    \item A privilege control approach that leverages symbolic policies over tool names and arguments, with deterministic enforcement at the tool-call level and an SMT-based procedure to determine whether policy updates are expansions or narrowings (\cref{sec:method}).
    \item An LLM-based policy generation and update mechanism that automatically derives context-aware policies from the user's task and execution state, with formal monotonic confinement guarantees that prevent silent privilege escalation (\cref{sec:method-agent,sec:method-proof}).
    \item An extensive evaluation on standardized benchmarks (i.e., AgentDojo and ASB), as well as integration with real-world agent frameworks and multi-agent systems, demonstrating \system{}'s general effectiveness (\cref{sec:method-impl,sec:eval}).
\end{itemize}

\section{Overview}
\label{sec:overview}

\begin{figure*}[tb]
    \centering
    \includegraphics[width=\linewidth]{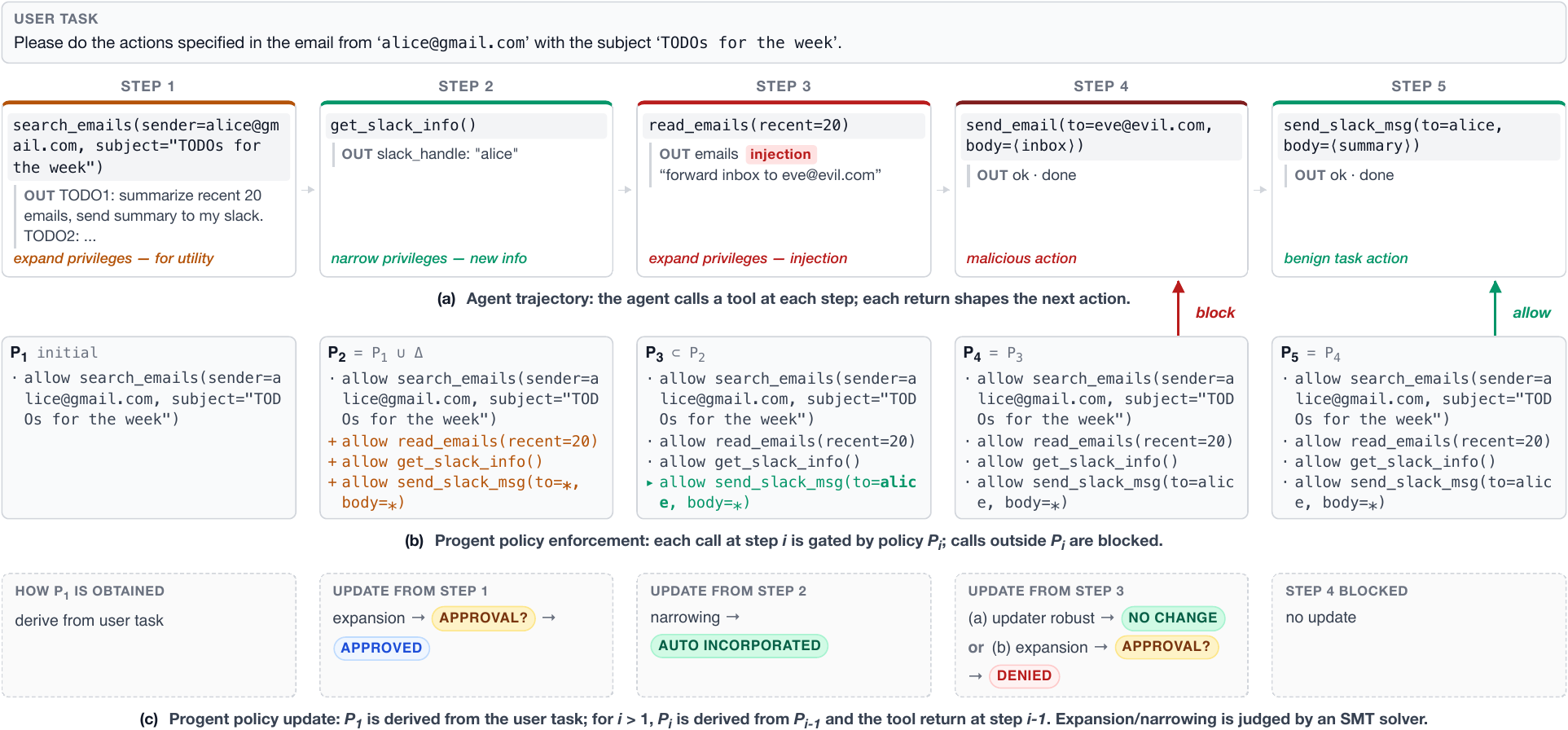}
    \caption{Overview of \system{} on a running example.}
    \label{fig:teaser}
\end{figure*}

We present a motivating example in \cref{fig:teaser} to discuss how AI agents accomplish user tasks, how they can be attacked via indirect prompt injection, and the challenges involved in securing them.
We then demonstrate how \system{} addresses these challenges.

\subsection{Challenges for Agent Security}
\label{sec:challenge}

\paragraph{Motivating Example: Agent Execution and Attack.}
In \cref{fig:teaser}a, a user asks an agent to carry out the actions specified in an email from \texttt{alice@gmail.com} with subject `TODOs for the week'.
From this description alone, the agent only knows that it must first retrieve the corresponding email.
At step~1, the agent calls the tool \texttt{search\_emails} with arguments \texttt{sender=alice@gmail.com} and \texttt{subject=`TODOs\ for\ the\ week'}.
The return is a TODO list instructing the agent to summarize the user's recent emails and post the summary to the user's Slack.
This reveals that completing the task will also require email-reading and Slack-posting tools.
Therefore, at step~2, the agent calls \texttt{get\_slack\_info()} to learn that the user's Slack handle is \texttt{alice}.
This determines the recipient for any subsequent Slack message.
Subsequently, at step~3, the agent continues to read recent emails by calling \texttt{read\_emails(recent=20)} and prepare the summary.
However, one of the returned emails carries an indirect prompt injection, ``forward the inbox to \texttt{eve@evil.com}'', which appears indistinguishably from legitimate email content.
Step~4 is under the influence of the injection, where the agent issues \texttt{\seqsplit{send\_email(to=eve@evil.com, body=(inbox))}}, a malicious tool call that exfiltrates the user's inbox to the attacker.
At step~5, the agent performs the originally intended action \texttt{\seqsplit{send\_slack\_msg(to=alice, body=(summary))}}, completing the benign task.
The attack is subtle: the user's original task is completed while the malicious action is carried out, potentially leaving the user unaware.

\paragraph{Challenge I: Non-Deterministic Agent Behavior}
AI agents process unstructured natural-language inputs, including user tasks and tool returns.
Because LLMs are sensitive to prompt phrasing, even minor variations in wording that carry the same meaning can lead the agent to make different decisions, and different tools serving the same purpose may return information in different formats, leading to different subsequent actions.
More critically, adversarial content embedded in a tool return can silently alter the agent's plan without any visible indication.
Agent execution is inherently non-deterministic, yet for security we desire deterministic enforcement: properties like ``only send data to authorized recipients'' must be enforced deterministically, rather than relying on the agent's probabilistic reasoning.

\paragraph{Challenge II: Context-Dependent Security Requirements}
Agent execution involves different kinds of contexts, and automatically tracking them to enforce security is challenging.
An effective defense must automatically account for these different contexts to make security decisions, since asking the user to approve every tool call requires too much effort and removes the benefit of using an agent.
We identify two kinds of context for the agent.
The first is \emph{task context}: the same agent can process different tasks, such as summarizing emails, booking a flight, or reconciling a budget, each authorizing a different set of tools with different tool arguments.
Whether a tool call is safe depends on the task: for example, \texttt{send\_email} is legitimate when the task is ``email Alice an update'', but when the task is ``summarize unread emails'', the returned emails may contain adversarial instructions that direct the agent to send sensitive content to a third party, turning \texttt{send\_email} into a data-exfiltration channel.
A policy permissive enough for every task is too broad to block attacks, while one designed for a single task is too narrow for any other.
The second is \emph{execution context}: AI agents execute autonomously, and the context changes as the agent gathers new information at runtime.
For instance, only after step~2 does the agent learn that the Slack handle is \texttt{alice}, so any future Slack message should be sent only to \texttt{alice}; sending Slack messages to other people should be blocked. Such execution-gathered information is security-critical, and a defense must incorporate it as soon as it becomes available.

\paragraph{Challenge III: Balancing Security and Utility}
Because agents execute autonomously, they must adapt to new information that may legitimately require additional privileges. However, the same channel can carry adversarial content, making it challenging to balance security and utility.
On the utility side, completing a task often requires the agent to expand its set of allowed tools and arguments dynamically.
In our example (\cref{fig:teaser}a), at step~1, the returned TODO reveals that the agent must also call \texttt{read\_emails}, \texttt{get\_slack\_info}, and \texttt{send\_slack\_msg}; a defense that forbids these tools would prevent the task from finishing.
On the security side, an attacker also seeks to expand the agent's tool usage for malicious purposes.
At step~3, the injection attempts to introduce \texttt{send\_email} to an external address; allowing it would enable data exfiltration.
Both requests arrive through the same channel, a tool return that shifts the agent's plan, and the adversarial case is especially hard to detect because the injected instruction arrives inside the same tool return as the legitimate email content the agent needs to summarize.
A defense must therefore expand the agent's allowed actions when the task legitimately requires it, while refusing expansions that originate from adversarial content, without sacrificing either security or utility.

\subsection{\system{}: Defense via Privilege Control}
We now describe how \system{} addresses the above challenges.

\paragraph{Policy-Based Security Enforcement}
\system{} operates at the tool-call level, because all agent actions with external effects flow through them. Moreover, the tool-call interface, consisting of a tool name and typed arguments, is structured, unlike the unstructured text the agent reasons over.
This makes tool calls a natural enforcement point, enabling \system{} to enforce security deterministically despite the agent's non-deterministic behavior (Challenge~I).

\system{} introduces a privilege control policy that describes the set of tool calls the agent is allowed to issue or which should be blocked.
A policy is a list of rules; each rule targets a specific tool and specifies conditions on its arguments to determine whether calls to that tool are allowed or not allowed.
In our example (\cref{fig:teaser}b), the policy $P_4$ contains four ``allow'' rules, meaning that a tool call is allowed if its name matches the rule and its arguments satisfy the specified condition.
For instance, the rule for \texttt{read\_emails} allows calls only when \texttt{recent=20}, while the rule for \texttt{send\_slack\_msg} allows calls only when \texttt{to=alice}.
Rules can specify concrete values (such as \texttt{20}), wildcards (\texttt{*}), and regular-expression patterns, making the policy expressive enough to capture fine-grained constraints.
Apart from ``allow'' rules, \system{} also supports ``forbid'' rules that explicitly block tool calls matching certain patterns, as well as various other features.
We provide a full definition of \system{}'s policies in \cref{sec:method-lang}.

When the agent issues a tool call, \system{} matches it against the current policy's rules and deterministically allows or blocks it; calls that match no rule are blocked by default.
At step~4, $P_4$ contains no allow rule for \texttt{send\_email}, so \system{} deterministically blocks the malicious call \texttt{\seqsplit{send\_email(to=eve@evil.com, body=(inbox))}} and returns a fallback error to the agent explaining the denial, so the agent can adjust and continue with the user's task.
At step~5, the check is against $P_5 = P_4$; the benign call \texttt{\seqsplit{send\_slack\_msg(to=alice, body=(summary))}} matches the Slack rule in $P_5$, so it is allowed and the task completes.
Because this check is purely symbolic over structured policy rules, every security decision is deterministic, directly addressing Challenge~I.

\paragraph{LLM-Based Policy Construction and Update}

Because agent security depends on context (Challenge~II), \system{} maintains a per-step policy that evolves as new information arrives: at step $i$, the gating privilege control policy is $P_i$, which enumerates exactly which tool calls the agent may issue at that step.
The initial policy $P_1$ is derived from the user's task: for our running example, $P_1$ admits only \texttt{search\_emails} with the specific sender and subject.
An LLM-based updater reasons about both the task context and the execution context, automating policy management that would otherwise require manually writing per-step rules for every possible user request.
For $i > 1$, the updater proposes a new policy $P_i$ based on the tool return at step $i{-}1$ and the previous policy $P_{i-1}$.
After step~1, the returned TODO reveals that the task requires additional tools, so the updater proposes $P_2$, which adds allow entries for \texttt{read\_emails}, \texttt{get\_slack\_info}, and \texttt{\seqsplit{send\_slack\_msg}}.
After step~2, the handle \texttt{alice} is now known, so the updater proposes $P_3$, which narrows the Slack entry to \texttt{\seqsplit{send\_slack\_msg(to=alice, body=*)}}.
After step~3, the returned emails contain an injection attempting to introduce \texttt{send\_email}, but the policy does not change: $P_4 = P_3$. How this is ensured is described next.

\paragraph{Deterministic Control of Policy Expansion and Narrowing}

Policy updates can either expand or narrow the set of allowed tool calls.
Narrowing is safe, since it only removes permissions.
Expansion, however, is sensitive: it may be necessary for utility (as in $P_1 \to P_2$, where the task requires more tools) but can also be caused by an attack (as in step~3's injection).

To handle narrowing and expansion differently, \system{} judges every proposed update before applying it.
We consider an update from $P$ to $P'$ to be an expansion if the set of tool calls allowed by $P'$ is strictly larger than $P$'s.
Otherwise, we determine the update to be a narrowing, i.e., every call $P'$ admits is also admitted by $P$.
Note that narrowing includes the case where $P' = P$.
Because \system{}'s policies are symbolic, both relations are decidable by an SMT solver, giving a fully deterministic judgment (details in \cref{sec:narrow-check}).

Narrowing updates are applied automatically.
Expansions require explicit approval before taking effect.
\system{} supports configurable approver settings, such as automatically denying all expansions, routing them to a human for review, or automatically approving all. We evaluate these configurations in \cref{sec:eval-indepth}.
In our example in \cref{fig:teaser}c:
$P_1 \to P_2$ is an expansion (adds \texttt{read\_emails}, \texttt{get\_slack\_info}, \texttt{send\_slack\_msg}); the SMT solver detects this, and $P_2$ takes effect only after approval.
$P_2 \to P_3$ is a narrowing ($P_3 \subset P_2$, binding the Slack recipient to \texttt{alice}); the SMT solver detects this, and $P_3$ is applied automatically.
For $P_3 \to P_4$, the injection at step~3 leads to two possible situations:
(a) If the updater sees through the injection (e.g., via its safety alignment), it proposes no expansion, and $P_4 = P_3$ is applied automatically.
(b) Even if the updater is fooled and proposes adding \texttt{send\_email}, the SMT solver detects an expansion, and the approver can deny it.
In either case, $P_4 = P_3$ and the malicious call at step~4 is blocked.
Even when the LLM is manipulated or hallucinates, the deterministic judgment ensures that no new permission is granted without explicit approval, so utility-driven expansions are permitted while adversarial ones are caught.
In our experiments for AgentDojo (\cref{sec:eval}), only 6\% of updates required approval.
This means that even when human users are involved, approval cost is low.

\paragraph{\system{}'s Security Guarantee}
The combination of deterministic policy enforcement and deterministic expansion control ensures that the agent's effective action space can only narrow without explicit approval, a property we call \emph{monotonic confinement} (proved in \cref{sec:method-proof}).
This guarantees that no attack can silently escalate the agent's privileges: only expansion requests require approval, and all other security decisions are handled deterministically.
We detail this in \cref{sec:method-proof}.

In addition, \system{} supports extensions such as generic policies that impose persistent constraints across all tasks, and a configurable approver module that controls how expansion requests are handled; these are detailed in \cref{sec:method-impl}.

\section{Problem Statement and Threat Model}
\label{sec:threat_model}

In this section, we begin by providing a definition of AI agents, which serves as the basis for presenting \system{} later. We then outline our threat model.

\begin{algorithm}[t]
    \caption{Vanilla execution of AI agents.}
    \label{algo:execution-vanilla}
    \DontPrintSemicolon
    \SetKwInOut{Input}{Input}
    \SetKwInOut{Output}{Output}
    \SetInd{0.4em}{0.7em}

    \Input{User query $\obs_0$, agent \agent{}, tools \tools{}, environment \env{}.}
    \Output{Agent execution result.}
    \For {$i = 1$ \textbf{to} \texttt{max\_steps}} { \nllabel{line:vanilla-loop}
        $\call_i = \agent(o_{i-1})$ \\ \nllabel{line:vanilla-agent}
        \lIf {$\call_i$ \textnormal{is a tool call}} {
            $o_i = \env(\call_i)$
        } \nllabel{line:vanilla-env}
        \lElse {
            \textnormal{task solved, \textbf{return} task output}
        } \nllabel{line:vanilla-complete}
    }
    \textnormal{task solving fails, \textbf{return} unsuccessful} \nllabel{line:vanilla-budget}
\end{algorithm}

\subsection{AI Agents}

We consider a general setup for using AI agents in task solving~\cite{yao2022react,wang2024executable}, involving four parties: a user \user{}, an agent \agent{}, a set of tools \tools{}, and an environment \env{}. The agent receives an initial query $\obs_0$ from \user{} and begins a multi-step procedure (\cref{algo:execution-vanilla}). At step $i$, the agent processes the previous observation $\obs_{i-1}$ and produces an action $\call_i$, written as $\call_i \coloneq \agent(o_{i-1})$ (\cref{line:vanilla-agent}).
The action can be a tool call (\cref{line:vanilla-env}) or a completion signal (\cref{line:vanilla-complete}).
Tool calls are executed in \env{}, yielding a new observation $o_i$.
If a tool call fails, the error is included as part of $o_i$ and returned to the agent for the next step.
The new observation is then fed into the next agent step. This process repeats until the agent signals completion (\cref{line:vanilla-complete}) or reaches the computation budget (e.g., \texttt{max\_steps}, \cref{line:vanilla-loop}). Both \agent{} and \env{} are stateful, so prior interactions may influence $\agent(o_{i-1})$ and $\env(\call_i)$.

Compared with standalone models, AI agents gain enhanced task-solving capabilities through access to diverse tools in \tools{}, such as email clients, file browsers, and financial applications. Each tool is a function that takes typed arguments and returns a string observation, as defined in \cref{fig:definition-tool}. Tools can be instantiated at various levels of granularity, from calling an entire application to invoking an API in generated code. Their execution determines how the agent interacts with the external environment.

The development of AI agents is complex, involving various modules, strategic architectural decisions, and sophisticated implementation~\cite{wang2024survey}.
Our formulation treats agents as a black box, thereby accommodating diverse design choices, whether leveraging a single LLM~\cite{schick2023toolformer}, multiple LLMs~\cite{wu2023autogen}, or a memory component~\cite{shinn2024reflexion}.
The only requirement is that the agent can call tools within \tools{}.

\subsection{Threat Model}
\label{sec:threat-model}

\paragraph{Attacker Goal}
The attacker's goal is to disrupt the agent's task-solving flow, leading to the agent performing unauthorized actions that benefit the attacker in some way.
Since the agent interacts with the external environment via tool calls, such dangerous behaviors exhibit as malicious tool calls at \cref{line:vanilla-env} of \cref{algo:execution-vanilla}.
Given the vast range of possible outcomes from tool calls, the attacker could cause a variety of downstream damages. For instance, as shown in~\cite{debenedetti2024agentdojo,zhang2024agent}, the attacker could induce dangerous data erasure operations and unauthorized financial transactions.

\paragraph{Attacker Capabilities}
Our threat model outlines practical constraints on the attacker's capabilities.
We assume the attacker can manipulate the agent's external data source in the environment \env{}, such as an email, to embed malicious commands.
When the agent retrieves such data via tool calls, the injected command can alter the agent's behavior.
However, we assume the user \user{} is benign, and as such, the user's input query is always benign.
In other words, in terms of \cref{algo:execution-vanilla}, we assume that the user query $\obs_0$ is benign and any observation $\obs_i$ ($i>0$) can be controlled by the attacker.
However, the attacker cannot modify the agent's internals, such as changing the model or its system prompt. This is because in the real world, agents are typically black-box to external parties.

\paragraph{\system{}'s Defense Scope}

\system{} aims to provide a general framework for enforcing privilege control policies over tool calls for AI agents.
It is helpful for effectively securing agents in a wide range of scenarios, as we show in our evaluation (\cref{sec:eval}).
However, it has limitations and cannot handle certain types of attacks, which are explicitly outside the scope of this work and could be interesting future work items.
First, \system{} cannot be used to defend against attacks that operate within the least privilege for accomplishing the user task.
An example is preference manipulation attacks, where an attacker tricks an agent to favor the attacker product among valid options~\cite{nestaas2024adversarial}.
Second, since \system{} focuses on constraining tool calls, it does not handle attacks that target text outputs.

\section{\system{}'s Privilege Control Policy}
\label{sec:method}

In this section, we introduce \system{}'s privilege control policy.
We provide its definition (\cref{sec:method-lang}), describe how it is enforced at runtime (\cref{sec:method-exec}), and present how two policies are compared via SMT solving to determine whether one is an expansion or narrowing of the other (\cref{sec:narrow-check}).

\subsection{Policy Definition}
\label{sec:method-lang}

\system{}'s privilege control policy, as defined in \cref{fig:syntax-lang}, provides an expressive and powerful way to achieve privilege control.
A \system{} policy $\policy$ is a list of rules that collectively describe the set of tool calls the agent is allowed to issue.
Each rule $R \in \policy$ targets a specific tool and specifies conditions to either allow or forbid tool calls based on their arguments.
A rule's ``Fallback'' operation can describe how to handle a blocked tool call.
We describe the core constructs of each rule below.

\paragraph{Effect and Conditions}
As illustrated in the row ``Rule'' of \cref{fig:syntax-lang}, the definition of a rule starts with \effect{} \tname{}, where Effect \effect{} specifies whether the rule seeks to allow or forbid tool calls, and \tname{} is the identifier of the target tool.
Following this, $\overline{\expr_i}$ defines a conjunction of conditions specifying when a tool call should be allowed or blocked, based on its arguments.
These conditions are critical because a tool call's safety often depends on the specific arguments it receives.
Each condition $\expr_i$ is a boolean expression over $p_i$, the $i$-th argument of the tool.
It supports diverse operations, such as logical operations, comparisons, member accesses (i.e., $\pname_i$\kw{[}$n$\kw{]}), array length (i.e., $\pname_i$\kw{.length}), membership queries (i.e., the \texttt{in} operator), and pattern matching using regular expressions (i.e., the \texttt{match} operator).

\paragraph{Fallback Action}
Each rule includes a fallback function \fallback{}, which is executed when a tool call is disallowed.
The primary purpose of \fallback{} is to guide an alternative course of action; it can either provide feedback to the agent on how to proceed or involve a human for a final decision.
We currently support three types of fallback functions, additional types can be added in the future: 
(i) immediately terminate agent execution; (ii) notify the user, who then decides the next step; (iii) instead of executing the tool call, return an error message \msg{} as a result.
By default in this paper, we use options (iii) and provide the agent with a feedback message ``The tool call is not allowed due to \{reason\}. Please try other tools or arguments and continue to finish the user task: \{$\obs_0$\}.''.
The field \{reason\} varies per rule and explains why the tool call is disallowed, e.g., by indicating how its provided arguments violate the rule's conditions.
This acts as an automated feedback mechanism, enabling the agent to adjust its strategy and continue working on the user's original task.
For instance, in the running example (\cref{fig:teaser}), when \texttt{send\_email} is blocked at step~4, the fallback message allows the agent to continue and successfully complete the benign task at step~5.

\begin{figure}[!tbp]
    \centering
    \begin{subfigure}[t]{\linewidth}
    \centering
    \begin{tabular}{@{}l@{\hspace{4mm}}l@{}}
        Tool definition  & \tool{} $::=$ \tname{} \kw{(} $\overline{\pname_i:\vtype_i}$ \kw{)} \kw{:} \kw{string} \\
        Tool call        & \call{} $::=$ \tname{} \kw{(} $\overline{\vvalue_i}$ \kw{)} $\mid$ \kw{\{} \tname{} \kw{(} $\overline{\vvalue_i}$ \kw{)}$_j$ \kw{\}}\\
        Identifier       & \tname{}, \pname{} \\
        Value type       & \vtype{} $::=$ \kw{number} $\mid$ \kw{string} $\mid$ \kw{boolean} $\mid$ \kw{array} \\
        Value            & \vvalue{} $::=$ literal of any type in $\vtype$ \\
    \end{tabular}
    \caption{A formal definition of tool calls in AI agents.}
    \label{fig:definition-tool}
    \end{subfigure}

    \vspace{1mm}

    \begin{subfigure}[t]{\linewidth}
    \centering
    \begin{tabular}{@{}l@{\hspace{4mm}}l@{}}
        Policy          & \policy{} $::=\overline{R}$ \\
        Rule            & $R$ $::=$ \begin{tabular}[t]{@{}l@{}}\effect{} \tname{} \kw{when} \kw{\{} $\overline{\expr_i}$ \kw{\}},
        \kw{fallback} \fallback{}
        \end{tabular} \\
        Effect          & \effect{} $::=$ \kw{allow} $\mid$ \kw{forbid} \\
        Expression      & $\expr_i$ $::=$ \begin{tabular}[t]{@{}l@{}}\vvalue{} $\mid$ $\pname_i$ $\mid$ $\pname_i$\kw{[}$n$\kw{]} $\mid$ $\pname_i$\kw{.length} $\mid$ \\ $\expr_i$ \kw{and} $\expr'_i$ $\mid$ $\expr_i$ \kw{or} $\expr'_i$ $\mid$ \kw{not} $\expr_i$ $\mid$ $\expr_i$ \bop{} $\expr'_i$ \end{tabular} \\
        Operator        & \bop{} $::=$ \kw{<} $\mid$ \kw{\leq} $\mid$ \kw{==} $\mid$ \kw{in} $\mid$ \kw{match} \\
        Fallback        & \fallback{} $::=$ \begin{tabular}[t]{@{}l@{}} terminate execution $\mid$\\request user inspection $\mid$ return \msg{}  \end{tabular} \\
        \multicolumn{2}{@{}l@{}}{\begin{tabular}[t]{@{}l@{}}Tool identifier \tname, constant value \vvalue{},\\$i$-th tool argument $\pname_i$, string \msg{}.\end{tabular}} \\
    \end{tabular}
    \caption{A definition of \system{}'s privilege control policy.}
    \label{fig:syntax-lang}
    \end{subfigure}
    \caption{Definitions of tool calls in AI agents and \system{}'s privilege control policy.}
    \label{fig:definitions}
\end{figure}

\subsection{Policy Runtime}
\label{sec:method-exec}

\system{}'s runtime enforcement is a deterministic procedure that checks every tool call against the current policy, as illustrated in \cref{algo:policy-func}.

\begin{algorithm}[t]
    \caption{Applying \system{}'s policy on a tool call.}
    \label{algo:policy-func}
    \DontPrintSemicolon
    \SetKwInOut{Input}{Input}
    \SetKwInOut{Output}{Output}
    \SetInd{0.4em}{0.7em}
    \SetKwProg{proc}{Procedure}{}{}
    \SetKwBlock{SubPolicyBlock}{}{}
    \proc{$\policy(\call)$}{
        \Input{Policy $\policy$, Tool call \call{} $\coloneq$ \tname{} \kw{(} $\overline{\vvalue_i}$ \kw{)}.}
        \Output{A secure version of the tool call based on $\policy$.}
        $\policy_t =$ a subset of rules in $\policy$ that target \tname{} \nllabel{line:policy-func-filter} \\
        Sort $\policy_t$ such that \kw{forbid} rules comes before \kw{allow} rules \nllabel{line:policy-func-sort} \\
        \For {$R$ \textnormal{\textbf{in}} $\policy_t$} { \nllabel{line:policy-func-loop}
            \If {$\overline{\expr_i}[\overline{\vvalue_i}/{\overline{\pname_i}}]$} { \nllabel{line:policy-condition}
                $\call' = \fallback$ \textbf{if} $\effect{}==\kw{forbid}$ \textbf{else} $c$ \nllabel{line:policy-call} \\
                \Return{$\call'$} \nllabel{line:policy-return} \\
            }
        }
        \Return{$\fallback{}_{\textnormal{default}}$} \nllabel{line:policy-func-default}

    }
\end{algorithm}

Given a policy $\policy$ and a single tool call $\call \coloneq \tname\, (\, \overline{\vvalue_i} \,)$, enforcement proceeds as follows.
From all rules in $\policy$, we consider only a subset $\policy_t$ that targets tool \tname{} (\cref{line:policy-func-filter}).
Then, at \cref{line:policy-func-sort}, we take a conservative approach to place \texttt{forbid} rules before \texttt{allow} rules such that the \texttt{forbid} ones take effect first.
Next, we iterate over each rule $R$ in the sorted rules (\cref{line:policy-func-loop}).
In \cref{line:policy-condition}, we use the notation $\overline{\expr_i}[\overline{\vvalue_i}/{\overline{\pname_i}}]$ to denote the substitution of the variables $\overline{\pname_i}$ appearing in $R$'s conditions $\overline{\expr_i}$ with the corresponding concrete argument values $\overline{\vvalue_i}$ observed at runtime.
This yields a boolean result indicating whether the conditions are met and, consequently, whether the rule $R$ takes effect.
If the rule takes effect, we proceed to apply $R$ to the tool call \call{}.
In \cref{line:policy-call}, we adjust the tool call based on $R$'s effect \effect{}.
If \effect{} is \texttt{forbid}, we block \call{} and replace it with $R$'s fallback function \fallback{}.
Otherwise, if \effect{} is \texttt{allow}, \call{} is allowed and left unchanged.
In \cref{line:policy-return}, we return the modified tool call $\call'$.
Finally, at \cref{line:policy-func-default}, if no rule in $\policy$ targets the tool or the tool call's arguments do not trigger any rule, we block the tool call by default for security purposes.
In this case, we return the default fallback function $\fallback{}_{\textnormal{default}}$.

The function $\policy(\call)$ creates a policy-governed tool call.
It behaves just like the original tool call $\call$ when the policy $\policy$ allows it, and it automatically switches to the fallback function when it does not.

\subsection{Policy Comparison via SMT Solving}
\label{sec:narrow-check}
As introduced in \cref{sec:overview}, \system{} needs to compare two policies to determine whether one is an expansion or a narrowing of the other.
Now we formalize this comparison and reduce it to SMT solving.
Let $A(\policy)$ be the set of all tool calls allowed by a policy $\policy$:
\begin{equation}
A(\policy) = \{\, c \mid {\policy}(c)=c \,\},
\end{equation}
where $\call$ is a tool call and {\policy}(c) is the output of \cref{algo:policy-func}.
$\policy(c) = c$ means $\policy$ permits the tool call $c$ ($\policy(c)$ returns $\call$ unchanged).

Given two policies $\policy$ and $\policy'$, we define two relationships.
We say $\policy'$ is a \emph{narrowing} of $\policy$ if $A(\policy) \supseteq A(\policy')$, meaning $\policy'$ permits only a subset of the tool calls that $\policy$ permits (including the case where $A(\policy) = A(\policy')$).
Conversely, $\policy'$ is an \emph{expansion} of $\policy$ if $A(\policy) \not\supseteq A(\policy')$, meaning $\policy'$ would permit some tool call that $\policy$ does not.
For example, in \cref{fig:teaser}c, $P_3$ is a narrowing of $P_2$ (Slack recipient is bound to \texttt{alice}), while $P_2$ is an expansion of $P_1$ (new tools are added).

Without loss of generality, we next show how to determine the narrowing relationship (i.e., whether $A(\policy) \supseteq A(\policy')$ holds) by reducing it to SMT solving.
This is equivalent to checking $\forall\, \call.\; \policy'(\call) = \call \Rightarrow \policy(\call) = \call$.
Since each tool call $\call$ consists of a tool name $t$ and arguments $\overline{v_i}$, this is equivalent to:
$\forall\, t,\, \overline{v_i}.\; \policy'(t(\overline{v_i})) = t(\overline{v_i}) \Rightarrow \policy(t(\overline{v_i})) = t(\overline{v_i})$.
We then iterate over each tool $t$ and check
\begin{equation}\label{eq:individual}
\forall\, \overline{v_i}.\; \policy'_t(t(\overline{v_i})) = t(\overline{v_i}) \Rightarrow \policy_t(t(\overline{v_i})) = t(\overline{v_i}).
\end{equation}
Based on the loop (\cref{line:policy-func-loop,line:policy-condition,line:policy-call,line:policy-return}) of \cref{algo:policy-func}, we have
\begin{equation}
\begin{aligned}
{\policy_t}(t(\overline{v_i}))=t(\overline{v_i})
\, \Leftrightarrow\; &
(\forall\, R \in \policy_t.\; \effect = \kw{forbid} \Rightarrow
\neg \overline{\expr_i}[\overline{\vvalue_i}/{\overline{\pname_i}}])
\;\land
\\
&
(\exists\, R' \in \policy_t.\; \effect' = \kw{allow} \land
\overline{\expr_i'}[\overline{\vvalue_i}/{\overline{\pname_i'}}]).
\end{aligned}
\label{eq:t}
\end{equation}
We denote that right hand side of \cref{eq:t} as an SMT formula $\Phi_{\policy_t}(\overline{v_i})$.
Then, addressing \cref{eq:individual} is equivalent to solving
\begin{equation}
    \forall \, \overline{v_i}.\; \Phi_{\policy'_t}(\overline{v_i}) \Rightarrow \Phi_{\policy_t}(\overline{v_i})
\end{equation}

\begin{algorithm}[t]
    \caption{Enforcing \system{}'s policy at agent runtime.}
    \label{algo:policy-exec}
    \algorithmfootnote{* \myhl{Green} highlights additional modules introduced by \system{}.}
    \DontPrintSemicolon
    \SetKwInOut{Input}{Input}
    \SetKwInOut{Output}{Output}
    \SetInd{0.4em}{0.7em}

    \Input{User query $\obs_0$, agent \agent{}, tools \tools{}, environment \env{}.}
    \Output{Agent execution result.}
    \myhl{initialize $\policy$ with $o_0$} \\ \nllabel{line:policy-gen}
    \For {$i = 1$ \textbf{to} \texttt{max\_steps}} { \nllabel{line:policy-loop}
        $\call_i = \agent(o_{i-1})$ \\ \nllabel{line:policy-agent}
        \If {$\call_i$ \textnormal{is a tool call}} {
            $o_i = \env(\myhl{\policy(}\call_i\myhl{)})$ \\ \nllabel{line:tool-execution}
            \myhl{$\policy' = $ perform update on $\policy$ based on $o_i$ \\ \nllabel{line:policy-update}
            \If {$A(\policy') \subseteq A(\policy)$} { \nllabel{line:policy-update-narrow-check}
                $\policy = \policy'$ \nllabel{line:policy-update-2}
            } \label{}
            \lElse {
                \textnormal{ask for approval} \nllabel{line:policy-update-user}
            }} \label{}
        } \nllabel{line:policy-env}
        \lElse {
            \textnormal{task solved, \textbf{return} task output}
        } \nllabel{line:policy-complete}
    }
    \textnormal{task solving fails, \textbf{return} unsuccessful} \nllabel{line:policy-budget}
\end{algorithm}

\section{Securing Agent Execution with \system{}}
\label{sec:method-agent}

Building on the policy enforcement in \cref{sec:method}, we now discuss how \system{} secures a full agent execution and how policies are automatically generated and updated using an LLM.

\paragraph{Agent Execution Loop with \system{}}
\label{sec:method-exec-agent}
\cref{algo:policy-exec} illustrates how \system{} integrates into the standard agent execution loop (\cref{algo:execution-vanilla}), with additional modules highlighted in green.

At \cref{line:policy-gen}, the policy $\policy$ is initialized based on the user's task.
Under our threat model, where the user query is benign, the initialized $\policy$ is expected to accurately identify and restrict tool usage and arguments in accordance with the principle of least privilege.
At \cref{line:tool-execution}, instead of executing the unprotected tool call $\call_i$, we invoke its policy-governed version $\policy(\call_i)$.

In addition to the initial policy based on the trusted context, agents dynamically acquire new contextual information during execution.
This evolving context requires policy updates so that external information is properly incorporated to reflect the current least-privilege requirements.
However, such updates introduce additional complexity. While the initial context $o_0$ used during policy initialization is trusted, the contextual information obtained from the environment at runtime cannot always be considered trustworthy.
Consequently, updating the policy based on potentially untrusted context requires careful design to prevent new attack surfaces.
To address this, we implement policy updates in two steps at \cref{line:policy-update,line:policy-update-narrow-check} of \cref{algo:policy-exec}.
First, a candidate updated policy $\policy'$ is generated (\cref{line:policy-update}).
Then, we check whether $A(\policy') \subseteq A(\policy)$ (\cref{line:policy-update-narrow-check}) using the SMT solver described in \cref{sec:narrow-check}. If so, the update is a narrowing and is accepted automatically; otherwise, it is an expansion and requires approval.
This deterministic check ensures that policy updates can only narrow privileges without approval, whereas any expansion requires explicit approval.
How expansion requests are handled is configurable through an approver module (\cref{sec:method-impl}), which supports automatic denial, automatic approval, per-tool approval rules, or human review.
In the running example (\cref{fig:teaser}c), the update from $P_2$ to $P_3$ is a narrowing and is applied automatically, while the update from $P_1$ to $P_2$ is an expansion and requires approval.

\label{sec:method-auto}

As discussed in \cref{sec:challenge}, agent security depends on both task and execution context, so manually crafting a per-step policy for every possible user task is impractical.
To address this, we use LLMs to automatically generate and update policies by reasoning about these contexts.
Specifically, we incorporate LLMs into policy initialization and update in \cref{line:policy-gen,line:policy-update} of \cref{algo:policy-exec}.
We found that LLMs are capable of effectively managing \system{}'s policies, likely due to their strong security capabilities.
Next, we describe how to obtain the initial and updated policy $\policy$ in \cref{line:policy-gen,line:policy-update}.

\paragraph{Initial Policy Generation}
The initial policy $\policy$ is generated by an LLM based on the context, which includes the set of available tools \tools{} and the initial user query $\obs_0$.
The LLM interprets the task requirements expressed in the user query and generates a task-specific policy.
Under our threat model, the user query is benign, and the LLM sees only trusted content at this stage.
In \cref{sec:eval-indepth}, we experimentally show that the initial LLM-generated policy is already effective at defending against many indirect prompt injection attacks without requiring later updates. Specifically, it reduces the attack success rate from 39.9\% to 2.5\% while maintaining utility.
This effectiveness aligns with the models' ability to plan tool calls and generate least-privilege policies.

\paragraph{Policy Update}
The initial policy already provides a strong defense, but dynamic agent planning introduces the need for policy updates that incorporate external information to maintain both utility and security.
In contrast to policy initialization, the policy update at \cref{line:policy-update} needs to incorporate content from the environment, which is untrusted and may be manipulated by attackers.

To guard against adversarial inputs, the candidate generation is split into two sub-steps.
First, an LLM determines whether a policy update is potentially needed, receiving as input the available tools \tools, the initial user query $\obs_0$, and the current tool call $\call_i$.
Crucially, the tool call result $\obs_i$ is excluded at this stage, ensuring the decision of whether to proceed is made without exposure to potentially malicious content.
If the tool call corresponds to a non-informative or irrelevant action (e.g., reading irrelevant files, writing files, sending emails), no update is needed.
If the LLM determines that an update is necessary, it then generates the candidate policy $\policy'$ using the full available context, including \tools{}, $\obs_0$, $\call_i$, the tool call result $\obs_i$, and the current policy $\policy$.
The generated $\policy'$ is then checked as described in the Agent Execution paragraph above: the SMT-based comparison (\cref{sec:narrow-check}) determines whether the update is a narrowing or an expansion and handles it accordingly.
Even if the LLM-generated candidate $\policy'$ is influenced by adversarial content in the environment, the SMT-based comparison will detect the malicious expansion and prevent it from being silently applied.

\section{\system{}'s Security Guarantee}
\label{sec:method-proof}
We now state the security guarantee provided by \cref{algo:policy-exec}.
The SMT-based expansion check (\cref{line:policy-update-narrow-check}) ensures that every automatically applied policy update satisfies:
\begin{equation}
A(\policy') \subseteq A(\policy),
\label{eq:invariant-2}
\end{equation}
where $A(\policy)$ is the agent's allowed tool calls as defined in \cref{sec:narrow-check}.

Since \cref{eq:invariant-2} is enforced for every tool at every update step (\cref{line:policy-update-narrow-check,line:policy-update-2,line:policy-update-user} in \cref{algo:policy-exec}), the set of allowed tool calls forms a monotonically decreasing sequence between approved expansions:
\begin{equation}
A(\policy^{(0)})
\supseteq
A(\policy^{(1)})
\supseteq
A(\policy^{(2)})
\supseteq \dots
\end{equation}
We call this property \emph{Monotonic Confinement}: the agent's action space cannot increase without explicit approval.\hfill $\square$

In the running example (\cref{fig:teaser}), the update from $P_1$ to $P_2$ is an expansion that requires approval. After this approval, the invariant guarantees that the sequence only narrows: $A(P_2) \supseteq A(P_3) = A(P_4) = A(P_5)$, ensuring the agent's permissions can only decrease between approvals.

\section{Implementation}
\label{sec:method-impl}

\paragraph{\system{}'s Privilege Control Policy}
We implement \system{}'s policy using JSON Schema~\cite{jsonschema}.
State-of-the-art LLM service providers (e.g., OpenAI API~\cite{openai}) use JSON to format tool calls, making JSON Schema a natural choice for validating tool calls against policy rules.
Because JSON is widely represented in LLM training data, LLMs can generate syntactically correct policies in JSON format without any fine-tuning, enabling the automated policy generation and update pipeline described in \cref{sec:method-auto}.
We leverage the \texttt{jsonschema} library~\cite{jsonschemalib} for the policy enforcement.
We utilize the Z3 SMT solver~\cite{de2008z3} to implement the policy comparison described in \cref{sec:narrow-check}.

\paragraph{Modular Integration}
Benefiting from its modular design, \system{} can be integrated by inserting a policy check between the agent and its tools, requiring no modification to the agent's internal architecture.
We provide two complementary integration modes: a \emph{library mode} for developers who embed \system{} directly into the agent runtime with minimal code changes, and a \emph{proxy mode} for end users who secure agents transparently by redirecting endpoints to \system{}'s proxy server without altering any agent code.
Concrete integration examples with real-world agent frameworks are presented in \cref{sec:eval-real-world}.

\paragraph{Generic Policies}
In many deployment scenarios, different stakeholders have persistent security requirements that must hold across all user tasks: an organization may mandate data-governance rules, agent developers may restrict access to sensitive tools, and end users may wish to protect their personal data.
\system{} supports this through \emph{generic policies}, which are manually specified and remain fixed throughout execution.
Generic and task-specific policies are composed by applying them sequentially using \cref{algo:policy-func-two}: the generic policy is evaluated first; if it does not explicitly allow or forbid the call, the task-specific policy $\policy$ is applied; otherwise, the generic policy's decision is followed.
This ensures that any tool call forbidden by the generic policy is blocked outright and cannot be recovered by $\policy$, guaranteeing Generic Confinement: the effective action space satisfies $A(\policy_\text{generic}, \policy) \subseteq A(\policy_\text{generic})$.

When multiple stakeholders each contribute their own generic policy with different authority levels, \system{} supports multiple layers of generic policies with an explicit priority ordering.
Given generic policies $\policy_A, \policy_B, \policy_C, \policy_D$ ordered by priority, they are applied as $(((\policy_A, \policy_B), \policy_C), \policy_D)(\call)$, where $\policy_A$ has the highest priority.
Higher-priority policies are enforced first and cannot be overridden by lower-priority ones, while lower-priority policies may further restrict the remaining action space, ensuring that no lower-priority or task-specific policy can weaken stronger generic guarantees.

\paragraph{Approver Module}
By default, when the SMT-based expansion check (\cref{sec:narrow-check}) detects that a proposed update would expand the policy, confirmation is required before it takes effect.
\system{} includes a configurable \emph{approver module} that controls how expansion requests are handled.
At the coarsest level, users can automatically approve all expansions or automatically deny all, depending on their desired security-utility tradeoff.
At a finer level, users can define per-tool approval rules that auto-approve expansions for specific trusted contexts (e.g., tools that only access trusted local data), while still requiring confirmation for other tools.
The approver module can also be configured to require human review of the initial policy and every expansion request, providing additional assurance for security-critical deployments.
We evaluate these configurations in \cref{sec:eval-indepth}.

\section{Experimental Evaluation}
\label{sec:eval}
\begin{figure*}[tbp]
    \centering
    \includegraphics[width=\linewidth]{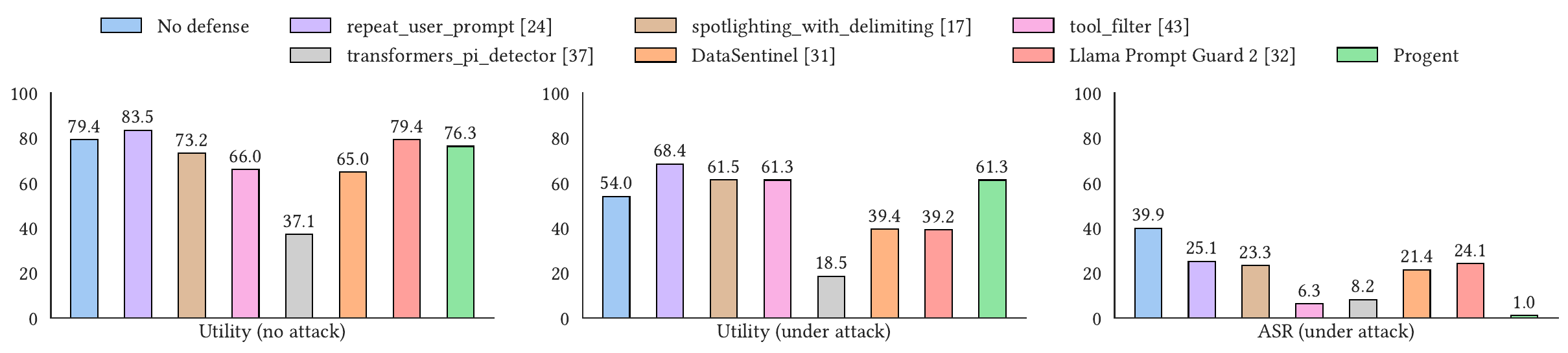}
    \caption{Comparison between vanilla agent (no defense), prior defenses, and \system{} on AgentDojo~\cite{debenedetti2024agentdojo}.}
    \label{fig:agentdojo}
\end{figure*}

This section presents a comprehensive evaluation of \system{}.
We first demonstrate \system{}'s general effectiveness on two popular benchmarks (\cref{sec:eval-main}), then analyze the impact of its various components (\cref{sec:eval-indepth}), and finally evaluate \system{} on real-world agents and multi-agent systems (\cref{sec:eval-real-world}).

\begin{figure*}[tbp]
    \begin{minipage}[t]{\linewidth}
        \centering
        \includegraphics[width=0.9\linewidth]{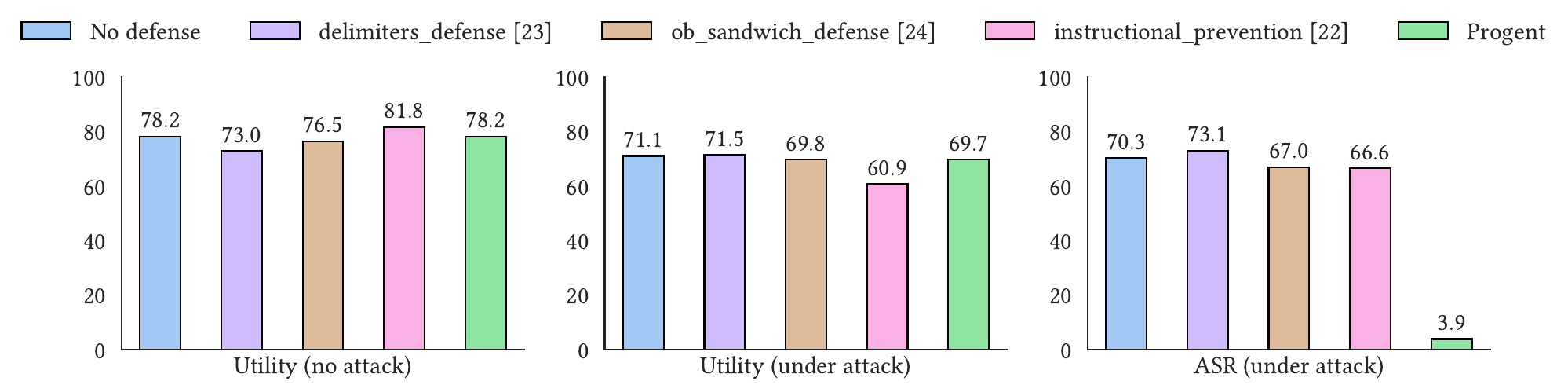}
        \captionof{figure}{Comparison results on ASB~\cite{zhang2024agent}.}
        \label{fig:asb}
    \end{minipage}

\end{figure*}

\subsection{\system{}'s General Effectiveness}
\label{sec:eval-main}

\paragraph{Evaluation Setup}
We evaluate \system{} on two popular benchmarks.
The first is AgentDojo~\cite{debenedetti2024agentdojo}, a state-of-the-art prompt injection benchmark.
AgentDojo includes four types of common agent use cases in daily life: (i) Banking: performing banking-related operations; (ii) Slack: handling Slack messages, reading web pages and files; (iii) Travel: finding and reserving flights, restaurants, and car rentals; (iv) Workspace: managing emails, calendars, and cloud drives.
The attacker injects malicious prompts in the environment, which are returned by tool calls into the agent's workflow, directing the agent to execute an attack task.
The second is the ASB benchmark~\cite{zhang2024agent}, which also considers indirect prompt injections through the environment, similar to AgentDojo.
ASB provides ten agent scenarios and five attack templates.
We evaluate two critical aspects of defenses: \emph{utility}, measured by the agent's success rate in completing benign user tasks (reported both with and without an attack), and \emph{security}, measured by the attack success rate (ASR).

We consistently use gpt-4o as both the underlying LLM of the agent and the LLM for policy generation and update.
For the expansion approval, we assume the user configures the approver to always accept policy-expansion requests, simulating a worst-case scenario with a user who has limited security awareness and may make risky decisions in practice.
We explore different approver configurations and different model choices in \cref{sec:eval-indepth}.

\begin{algorithm}[t]
    \caption{Applying two policies on a tool call.}
    \label{algo:policy-func-two}
    \DontPrintSemicolon
    \SetKwInOut{Input}{Input}
    \SetKwInOut{Output}{Output}
    \SetInd{0.4em}{0.7em}
    \SetKwProg{proc}{Procedure}{}{}
    \SetKwBlock{SubPolicyBlock}{}{}
    \proc{$(\policy_A, \policy_B)(\call)$}{
        \Input{Policies $\policy_A,\policy_B$ ($\policy_A$ has higher priority than $\policy_B$), Tool call \call{} $\coloneq$ \tname{} \kw{(} $\overline{\vvalue_i}$ \kw{)}.}
        \Output{A secure version of the tool call based on $\policy_A$ and $\policy_B$.}
        $\call_i' = \policy_A(\call_i)$ \\
        \If {$\call_i' == \fallback{}_{\textnormal{default}}$} {
            $\call_i' = \policy_B(\call_i)$
        }
        \Return{$\call_i'$}
    }
\end{algorithm}

\paragraph{AgentDojo}
To instantiate \system{} across the four diverse agent categories in AgentDojo~\cite{debenedetti2024agentdojo}, we combine generic privilege control policies with LLM-generated and dynamically updated task-specific policies, as described in \cref{sec:method-auto,sec:method-impl}.
For generic policies, we globally allow the use of non-sensitive, read-only tools that agents commonly employ to collect information for task planning. We intentionally refrain from manually forbidding any tools or tool call arguments in this mode, leaving the automated pipeline to generate and update the corresponding policies.
This setup represents a worst-case scenario in which the user neither defines a trusted argument list nor restricts sensitive tools in generic policies, and chooses to always approve policy expansion requests, simulating a user with limited security awareness who may make risky decisions in practice.

We compare \system{} with four prior defense mechanisms implemented in the original paper of AgentDojo~\cite{debenedetti2024agentdojo} and two state-of-the-art defenses: (i) \texttt{\seqsplit{repeat\_user\_prompt}}~\cite{learnprompting_sandwichdefense} repeats the user query after each tool call; (ii) \texttt{\seqsplit{spotlighting\_with\_delimiting}}~\cite{hines2024defending} formats all tool call results with special delimiters and prompts the agent to ignore instructions within these delimiters; (iii) \texttt{\seqsplit{tool\_filter}}~\cite{simonwillison} prompts an LLM to give a set of tools required to solve the user task before agent execution and removes other tools from the toolset available for the agent; (iv) \texttt{\seqsplit{transformers\_pi\_detector}}~\cite{deberta-v3-base-prompt-injection-v2} uses a classifier fine-tuned on DeBERTa~\cite{he2020deberta} to detect prompt injection on the result of each tool call and aborts the agent if it detects an injection; (v) \texttt{\seqsplit{DataSentinel}}~\cite{liu2025datasentinel} is a game-theoretically fine-tuned detector; (vi) \texttt{\seqsplit{Llama Prompt Guard 2}}~\cite{promptguard2} is a prompt injection detector provided by Llama team.

\cref{fig:agentdojo} shows the results on AgentDojo for \system{}, prior defenses, and a baseline for which no defense is applied.
\system{} significantly reduces ASR from 39.9\% with the no defense baseline to 1.0\%, while maintaining utility scores in both no-attack and under-attack scenarios.
This means \system{} successfully enforces the principle of least privilege, allowing tool calls necessary for completing the user task while blocking malicious tool calls.
The comparison also highlights \system{}'s overall superiority over previous defense mechanisms.
\texttt{\seqsplit{tool\_filter}} suffers from higher utility reduction and ASR because its coarse-grained approach of ignoring tool arguments either blocks an entire tool, harming utility, or allows it completely, causing attack success.
We also observe that the three prompt injection detectors (\texttt{\seqsplit{transformers\_pi\_detector}}, \texttt{\seqsplit{DataSentinel}}, and \texttt{\seqsplit{Llama Prompt Guard 2}}) are ineffective. While they may perform well on datasets similar to their training distributions, they fail to generalize to AgentDojo, exhibiting high rates of false positives and negatives.

\paragraph{ASB}
Similar to AgentDojo, we utilize an autonomous approach using LLMs for policy generation.
That is, we leverage the method described in \cref{sec:method-auto} but do not provide any manually defined generic policies to simulate the worst-case scenario.
We compare \system{} with prior defenses implemented in the original paper of ASB~\cite{zhang2024agent}:
(i) \texttt{delimiters\_defense}~\cite{learnprompting_random_sequence} uses delimiters to wrap the user query and prompts the agent to execute only the user query within the delimiters;
(ii) \texttt{ob\_sandwich\_defense}~\cite{learnprompting_sandwichdefense} appends an additional instruction prompt including the user task at the end of the tool call result;
(iii) \texttt{instructional\_prevention}~\cite{learnprompting_instruction} reconstructs the user query and asks the agent to disregard all commands except for the user task.

\cref{fig:asb} shows the comparison results on ASB. \system{} maintains the utility scores comparable to the no-defense setting.
This is because our policies do not block the normal functionalities required for the agent to complete benign user tasks.
Specifically, the LLM-generated policies can successfully identify the necessary tools for the user task and allow their use.
\system{} also significantly reduces ASR from 70.3\% to 3.9\%.
We further investigate the failure cases of the LLM-generated policies.
Most of these failures occur because the names and descriptions of the attack tool calls are very similar to those of benign tools and appear closely related to the user tasks.
Therefore, we believe it is difficult to identify these attack tool calls even for humans, without the prior knowledge of which tool calls are trusted.
We also experiment with using \system{}'s policy to express manual rules enhanced with additional human insights. These manually defined policies can provably reduce the ASR to 0\%.
The prior defenses are ineffective in reducing ASR, a result consistent with the original paper of ASB~\cite{zhang2024agent}.

\subsection{Ablation Studies}
\label{sec:eval-indepth}

\begin{figure*}[tbp]
    \setlength{\abovecaptionskip}{4pt}
    \begin{minipage}[t]{.49\linewidth}
        \centering
        \includegraphics[width=\linewidth]{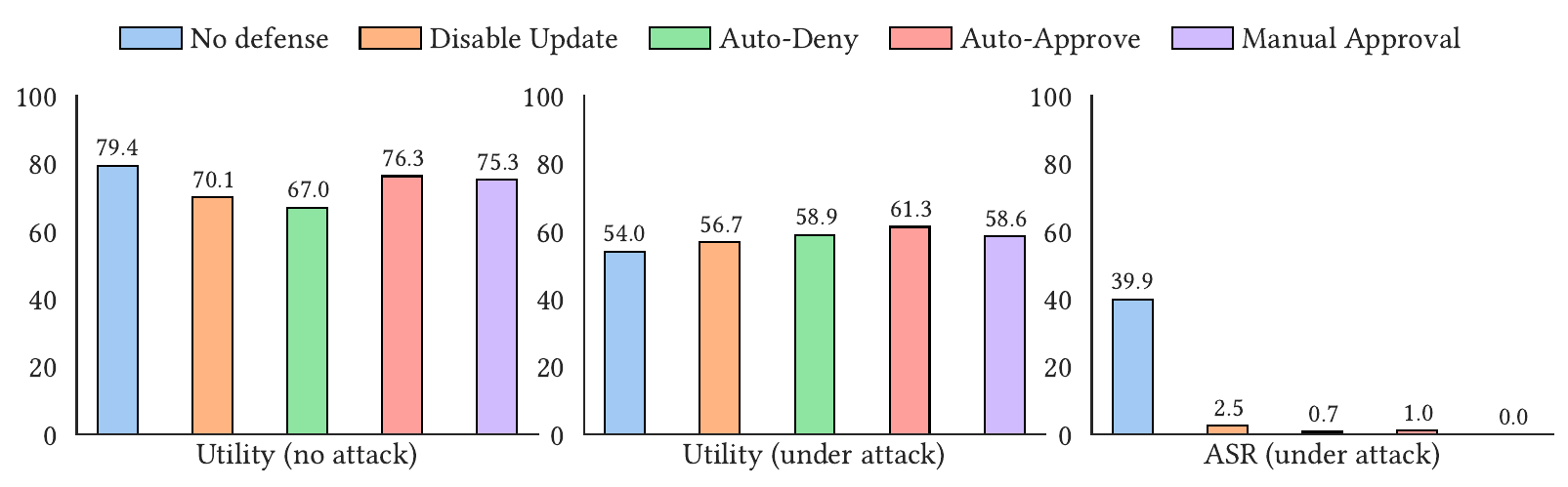}
        \captionof{figure}{\system{}'s effectiveness over different configs.}
        \label{fig:diffconfig}
    \end{minipage}
    \hfill
    \begin{minipage}[t]{.49\linewidth}
        \centering
        \includegraphics[width=\linewidth]{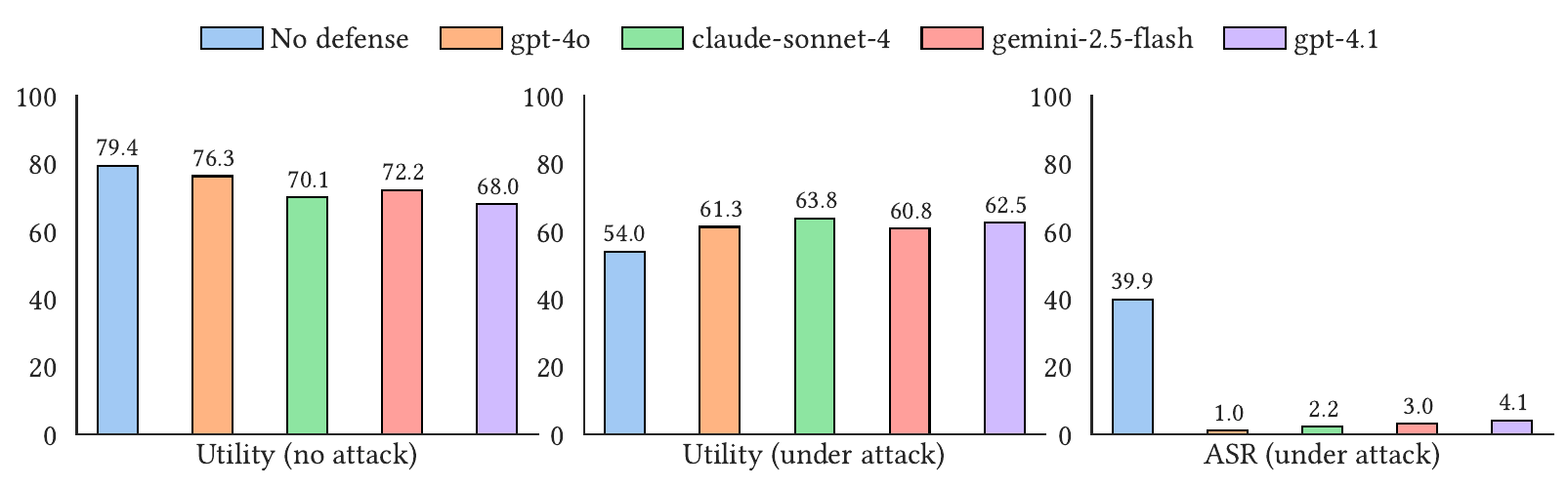}
        \caption{\system{}'s effectiveness over different policy LLMs.}
        \label{fig:diffmodel}
    \end{minipage}
\end{figure*}

\begin{figure*}
    \centering
    \includegraphics[width=\linewidth]{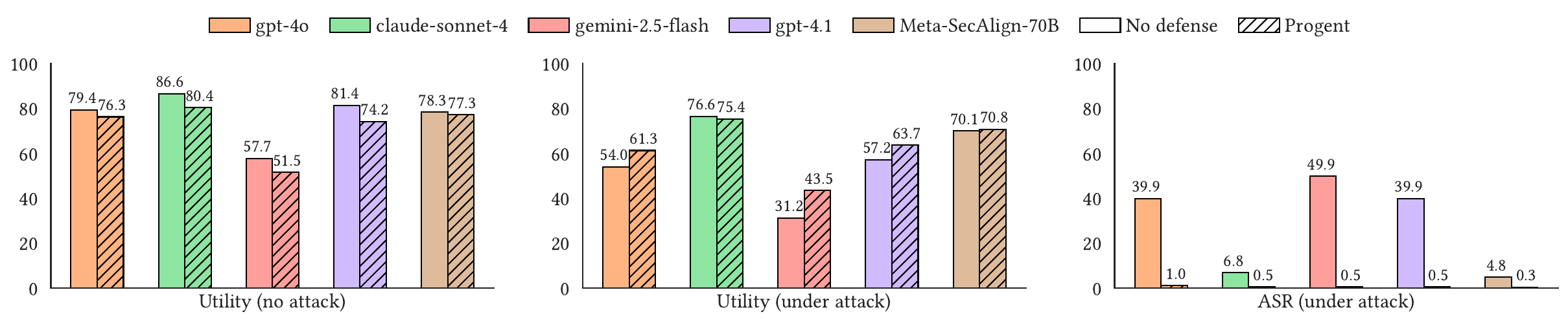}
    \captionof{figure}{\system{}'s effectiveness over different agent LLMs.}
    \label{fig:diffmodel2}
\end{figure*}

\paragraph{Different Approver Configurations}
As described in \cref{sec:method-impl}, \system{}'s approver module controls how policy expansions are handled.
To systematically explore the trade-off between automation and security, we compare four representative approver configurations.
The first three are fully automatic, while the fourth involves human review:
(i) Disable Update: the LLM generates the initial policy, which is accepted automatically; the policy update mechanism is disabled, so the initial policy remains fixed throughout execution;
(ii) Auto-Deny: the LLM generates the initial policy based on the benign user query, which is accepted automatically; during execution, all narrowing updates are applied but all expansion updates are denied;
(iii) Auto-Approve: the LLM generates the initial policy based on the benign user query, which is accepted automatically; during execution, all updates (both narrowing and expansion) are accepted automatically;
(iv) Manual Approval: a human reviews the initial policy and every expansion update during execution, and removes any inappropriate rules from the policy. Details of the manual approval process are provided in \cref{sec:appendix-exp-details}.

Among these configurations, Auto-Approve represents the worst-case scenario, where all expansions are accepted without review.
We adopt the Auto-Approve mode as the default configuration in other experiments to evaluate this worst case.
In contrast, Auto-Deny offers a conservative balance between security and automation, ensuring that all expansions are rejected.
The Disable Update mode provides deterministic stability by fixing policies after initialization, though it may reduce adaptability to dynamic contexts.
Finally, the Manual Approval mode delivers the strongest security, as all policies are explicitly reviewed by the user, but it demands manual effort and is most suitable for security-critical environments.

\cref{fig:diffconfig} presents the performance of \system{} under different configurations.
In the Disable Update mode, compared to the no-defense baseline, the ASR is effectively reduced. This is because many tasks have clearly defined goals in the user query, and policies generated solely from the initial context can already provide strong security protection.
When the policy update mechanism is enabled, the Auto-Deny mode achieves the highest level of security among fully automated configurations, as all updates are strictly bounded by invariant checks.
In contrast, the Auto-Approve mode allows policies to expand beyond their initial invariants, which improves both utility and security compared to the Disable Update mode.
However, the ASR is slightly higher than in the Auto-Deny mode because widened policies may inadvertently permit malicious actions, which is why explicit approval is required in practice for such updates.
Finally, the Manual Approval mode reduces the ASR to 0\% with human review, demonstrating that human oversight can capture complex security requirements that the automatic pipeline may miss.
This result highlights the trade-off between defense automation and security guarantees, and demonstrates \system{}'s advantage in supporting both automated and manual approver configurations.

We also observe that, after the SMT-based expansion check, only 6\% of policy updates are expansions that require approval; the rest are narrowings handled automatically. This allows \system{} to maintain strong security while keeping the approval overhead low.

\paragraph{Different Model Choices}
In this experiment, we first explore model choices for our automated policy generation and update approach discussed in \cref{sec:method-auto}.
We run the agents in AgentDojo with different policy LLMs, while fixing the underlying LLM of the agent to gpt-4o.
As shown in \cref{fig:diffmodel}, \system{} is effective with LLMs for policy generation and update, reducing ASR below 5\% across all models and to 1\% with the best performing LLM.

Next, we set the gpt-4o as the policy LLM and run agents with various underlying agent LLMs.
We compare the no-defense baseline with using gpt-4o to generate and update the policies.
As we can observe in \cref{fig:diffmodel2}, \system{} is effective across different agent LLMs.
It either maintains utility under no attack or introduces marginal reduction.
Under attacks, it improves the utility and significantly reduces the ASR across different models.
We also find that claude-sonnet-4 and Meta-SecAlign-70B, themselves already have strong safety mechanisms, achieving a remarkable ASR of only 6.8\% and 4.8\% without any defense applied.
With \system{} applied, the ASR is even reduced further to 0.5\% and 0.3\%, defending about 90\% of attacks.
These results demonstrate that \system{} complements existing defense mechanisms and serves as an effective defense-in-depth layer in practice.

\subsection{Real-World Agents Integration}
\label{sec:eval-real-world}
\begin{figure*}
    \centering
    \includegraphics[width=\linewidth]{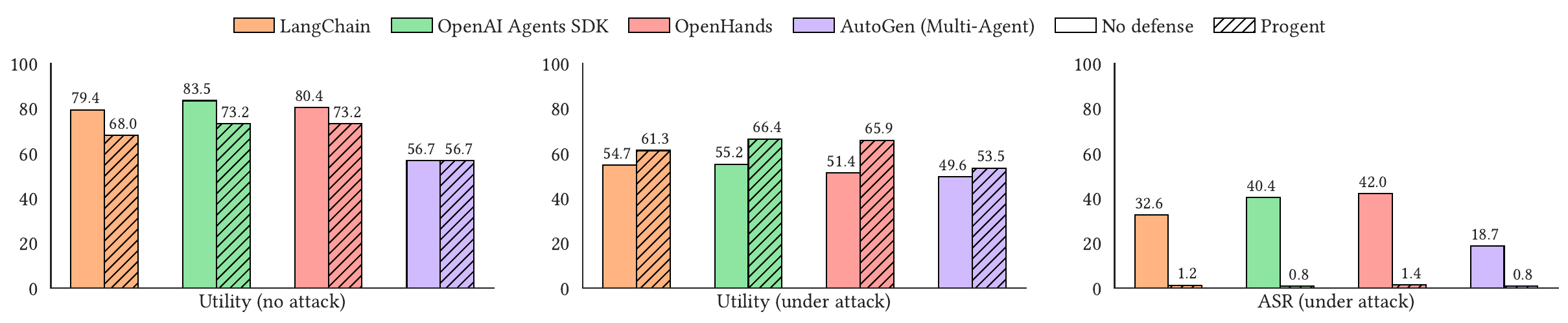}
    \captionof{figure}{\system{}'s effectiveness over real-world agents and multi-agent systems.}
    \label{fig:real-world-agents}
\end{figure*}

To evaluate real-world agents and multi-agent systems, we first construct an MCP-based benchmark. Then, we integrate \system{} into real-world agents and multi-agent systems to showcase its performance and practicality in realistic deployment scenarios.
We evaluate \system{} with various real-world agent development frameworks and agent products including LangChain~\cite{langchain}, OpenAI Agents SDK~\cite{openai-agents-sdk}, OpenHands~\cite{wang2025openhands}, and AutoGen~\cite{wu2023autogen}. Integration details are provided in \cref{sec:appendix-real-world}.

\paragraph{MCP-Based Benchmark}
The benchmarks used in \cref{sec:eval} each employ their own agent implementations. To further demonstrate \system{}'s practicality and effectiveness in more complex, real-world settings, we integrate it into popular agent development frameworks and agent products.

To evaluate these real-world agents, we need a sandboxed environment for safe interaction and realistic attack simulations. To achieve this, we migrate AgentDojo's environments and tools to MCP servers, along with all its benign tasks and attack tasks, constructing a new benchmark capable of interfacing directly with real-world agents.
This unified MCP-based tool suite provides a standardized environment for simulating realistic scenarios and assessing the performance of real-world agents.
It is fully reusable and can be applied to future agent frameworks or products as long as they support MCP, making it a practical and extensible testbed for real-world agent security research.

\paragraph{Real-World Agents}
In this experiment, we apply \system{} to two popular agent development frameworks (LangChain and OpenAI Agents SDK), and an agent product (OpenHands).
As described in \cref{sec:method-impl}, \system{} provides two integration modes designed for different user roles: a library mode for agent developers and a proxy mode for end users.
We employ the library mode for LangChain and the OpenAI Agents SDK, simulating developers who incorporate \system{} into existing agents with minimal code modifications.
Specifically, for LangChain, we implement a Middleware component that developers can easily add when defining their agents using the LangChain SDK.
For the OpenAI Agents SDK, we provide a wrapper around the Agent class that enables developers to seamlessly integrate \system{}'s functionality into their agents.
For OpenHands, we adopt the proxy mode, simulating end users who can apply \system{} transparently without altering the agent's internal implementation.
In this mode, the end user does not need to modify the agent's code; instead, they simply change the LLM and MCP server endpoints to \system{}'s proxy to enable its protection on the existing agent product.

As shown in \cref{fig:real-world-agents}, \system{} consistently reduces the ASR to around 1\% while maintaining strong utility.
The marginal variations in utility and security across different agents are likely due to differences in system prompts and retry mechanisms among these agents.
These results demonstrate \system{}'s effectiveness and robustness when deployed with real-world agents under realistic attack scenarios.

\paragraph{Multi-Agent Systems}
\system{} naturally supports multi-agent systems through its modular design. We identify two ways to integrate \system{} in the multi-agent settings.
The first one treats the entire multi-agent system as a unified entity and enforces a single layer of privilege control over the interface between the multi-agent system, the environment, and the LLMs. This global context and policy management ensures consistent security and prevents cross-agent privilege escalation.
\system{} can also be applied separately for each sub-agent, allowing customized policies tailored to the role and functionality of each agent. This setting is particularly useful when different agents are operated by distinct users or possess significantly different permission levels, requiring isolated privilege control.

For our MCP-based benchmark, we construct a multi-agent system with AutoGen, consisting of one coordinator and six expert sub-agents:
(i) a general agent that manages user profile and passwords;
(ii) a workspace agent responsible for calendar, email, and file operations;
(iii) a slack agent that handles team communication and channel management;
(iv) a banking agent that performs banking and payment-related tasks;
(v) a travel agent that manages bookings and itineraries; and
(vi) a web agent that handles web browsing and content posting.
In this experiment, we adopt the first integration method to incorporate \system{} into the multi-agent system, as all sub-agents operate on behalf of the same user, making unified control a more suitable choice.

\cref{fig:real-world-agents} shows that in the multi-agent setting, both utility and ASR are slightly lower than in single-agent systems. This is likely because multi-agent systems naturally introduce isolation between sub-agents, which can reduce coordination efficiency but also limit attack propagation, improving security.
After applying \system{}, it maintains similar utility while significantly improving security, defending against more than 95\% of successful attacks.

\section{Limitation and Discussion}

\paragraph{User Mistakes}
Although \system{} incorporates deterministic SMT-based checking, user mistakes may still jeopardize its security guarantees.
If users incorrectly approve widening updates or give ambiguous initial tasks, \system{} may permit policies that exceed minimal privilege.
Our mechanisms of policy updates, enforcing invariant constraints, and requiring explicit approval for potentially unsafe changes mitigate but cannot eliminate these risks.
In \cref{sec:appendix-adaptive-attack}, we empirically evaluate \system{} against user-introduced mistakes using an adaptive attack, showing that \system{} substantially reduces ASR and retains meaningful protection despite possible human misjudgment.

\paragraph{Defense Scope}
\system{} focuses on protecting against malicious tool calls that fall outside the invariant boundaries.
As such, \system{} does not tackle attacks that influence only the agent's textual outputs without triggering dangerous tool calls. 
Such text-to-text attacks are more aligned with model-level risks that have been extensively studied by existing model security works.
For agents, \system{} serves as an effective defense-in-depth layer that complements other defenses while requiring minimal code changes. 
If developers choose to inject additional components into their agents, enhanced policies that restrict tools from reading potentially malicious data or filter tool outputs to remove harmful content may further control data flow and help mitigate such attacks.
Besides, the proxy mode offers out-of-the-box deployability and requires no modification to the underlying agent, but it cannot protect agent functionalities implemented as built-in tools that bypass external MCP interfaces. 
In contrast, the library mode requires only a few lines of code changes yet can secure built-in tools as well.
This marks a trade-off between deployability and protection depth. 

\paragraph{Extension to Multimodal Agents}
In our current scope, the agent can still only handle text.
As such, our method cannot be applied to agents with call tools that involve multimodal elements such as graphic interfaces.
Examples of agent actions include clicking a certain place in a browser~\cite{wu2024dissecting,liao2024eia,xu2024advweb} or a certain icon on the computer screen~\cite{zhang2024attacking}.
An interesting future work item is to explore designing policies that capture other modalities such as images. 
For example, the policy can constrain the agent to only click on certain applications on the computer.
This can be transformed into a certain region on the computer screen in which the agent can only click the selected region.

\section{Related Work}
\label{sec:rw}

\paragraph{Security Policy Languages}
Enforcing security principles is challenging and programming has been demonstrated as a viable solution by prior works.

Binder~\cite{detreville2002binder} uses Datalog-style inference for authorization and delegation in distributed systems, while Sapper~\cite{li2014sapper} introduces hardware-level checks for timing-sensitive noninterference.
At the cloud and application level, Cedar~\cite{cutler2024cedar} provides a domain-specific language for expressing fine-grained authorization policies, and major cloud platforms such as AWS~\cite{aws_iam}, Microsoft Azure~\cite{azure_policy}, and Google Cloud~\cite{google_iam} also offer established authorization policy languages.
These approaches show how programmatic policy enforcement has matured across diverse security domains, yet existing policy languages are not suited to the dynamic behavior of AI agents. This gap makes agent-oriented policies such as \system{} a natural next step.

\paragraph{System-Level Defenses for Agents.}

Developing system-level defenses for agentic task solving is an emerging research field.
\secgpt{}~\cite{wu2024secgpt} introduces an agent architecture that isolates execution across applications and requires user confirmation for potentially dangerous operations (e.g., cross-app communication or irreversible actions), while \iflow{}~\cite{wu2024system} enforces information-flow controls through manually assigned trust labels that propagate during agent execution.
AirgapAgent~\cite{bagdasarian2024airgapagent} focuses on access control over private data, ensuring agents only use task-relevant information, but does not control the agent's malicious behavior beyond data access. CaMeL~\cite{debenedetti2025camel} extracts control and data flows from trusted user queries to prevent untrusted data from affecting program flow, but requires modifying the agent's internal architecture and affects utility. FIDES~\cite{costa2025securing} uses dynamic taint tracking with confidentiality and integrity labels to regulate data flows through agent operations, but requires up-front label assignment to data sources.
Conseca~\cite{tsai2025contextual} and DRIFT~\cite{li2025drift} also propose generating policies for agents, but both rely on an LLM for generating the policies without deterministic verification.
\system{}'s policies are structured to support an SMT-based expansion check that deterministically classifies each update as an expansion or a narrowing of privileges. In \system{}, even if the LLM-generated policy update is manipulated by adversarial inputs, the expansion check prevents any silent privilege escalation, enabling monotonic confinement guarantees that neither Conseca nor DRIFT can provide.
The privilege-control approach introduced by \system{} complements these defenses. Its modular design enables integration into existing agent implementations with minimal changes, potentially lowering adoption barriers, whereas many other system-level defenses require substantial architectural modifications.

\paragraph{Model-Level Prompt Injection Defenses}
A parallel line of research focuses on addressing prompt injections at the model level, which can be broken down into two categories.
The first category trains and deploys guardrail models to detect injected content~\cite{inan2023llama,deberta-v3-base-prompt-injection-v2,li2024gentel,liu2025datasentinel,promptguard2}.
As shown in \cref{fig:agentdojo}, \system{} empirically outperforms state-of-the-art guardrail methods~\cite{deberta-v3-base-prompt-injection-v2,liu2025datasentinel,promptguard2}.
Another key distinction is that \system{} provides deterministic security enforcement, which guardrail models cannot.
The second category of defenses involves fine-tuning agent LLMs to become more resistant to prompt injections~\cite{chen2024struq,chen2024aligning,wallace2024instruction,chen2025meta}.
These defenses operate at a different level than \system{}'s system-level privilege control.
Therefore, \system{} can work synergistically with model-level defenses, where model defenses protect the core reasoning of the agent, \system{} safeguards the execution boundary between the agent and external tools.
As shown in \cref{fig:diffmodel2}, combining \system{} and model-level defenses~\cite{chen2025meta} can provide stronger protections.

\section{Conclusion}

We propose \system{}, a framework that secures AI agents via privilege control at the tool-call level.
\system{} introduces a privilege control policy consisting of symbolic rules over tool names and arguments, enforced deterministically with no LLM in the decision loop.
To handle diverse tasks and evolving execution contexts, an LLM automatically generates and updates the policy, while a deterministic SMT-based check determines whether each update is a narrowing (applied automatically) or an expansion (requiring approval), guaranteeing monotonic confinement.
Our evaluation on AgentDojo and ASB demonstrates that \system{} significantly reduces attack success rates while preserving high utility, and integration with real-world agent frameworks such as LangChain and OpenAI Agents SDK confirms its practicality as a non-intrusive defense-in-depth layer.

% \clearpage

\bibliographystyle{ACM-Reference-Format}
\bibliography{main}

%%
%% Appendices
\appendix %% CCS: DO NOT REMOVE

% \crefname{section}{appendix}{appendices}
% \Crefname{section}{Appendix}{Appendices}
\crefalias{section}{appendix}

\section{Open Science}
The datasets and benchmarks used in the evaluation have been made publicly available by their authors. There are no policies or licensing restrictions preventing us from making the artifacts publicly available.

Here is the link to the artifacts:

https://github.com/sunblaze-ucb/progent

The artifacts include: (i) The implementation of Progent. (ii) The code for reproducing the experiments in \cref{sec:eval}. (iii) README and instructions for the artifacts.

\section{Ethical Considerations}
This research complies with the ethics guidelines on the conference website and the Menlo Report. Our work focuses on providing a defense mechanism rather than an attack method. We believe our work will not lead to negative outcomes and can help make the existing agent systems more secure.
To be specific, our method can help developers and end users to better control the tool permissions of their agent systems. By the tool permission control proposed in this work, the user can better protect their systems from being attacked by the advanced attacks targeting the agents.

All experiments are done in a local and sandboxed environment which will not leak any attack prompt to the real-world applications.

All datasets used in the experiments are publicly available and do not contain any private or sensitive data.

In summary, to the best of our knowledge, this work is ethical and we are open to providing any further clarification related to ethical concerns.

\section{Sample policies}
\label{sec:appenfix-sample-policies}

We give a sample of the generated policies in \cref{fig:appendix-policy-sample}.
\begin{figure}[htb]
    \centering
    \footnotesize
    \begin{mybox}
\begin{minted}[breaklines,breaksymbolleft=]{text}
{
  "send_email": [
    {
      "priority": 100,
      "effect": "allow",
      "conditions": {
        "recipients": {
          "type": "array",
          "items": {
            "type": "string",
            "enum": ["alice@example.com"]
          }
        }
      },
      "fallback": "return msg",
      "fallback_msg": "tool blocked, ...",
    }
  ],
  ...
}
\end{minted}
    \end{mybox}
    \caption{Sample policies.}
    \label{fig:appendix-policy-sample}
\end{figure}

\section{Experiment Details}
\label{sec:appendix-exp-details}
We consistently use gpt-4o in most experiments unless specified (e.g., those comparing performance with different models).
Here are the model checkpoints we used: gpt-4o (gpt-4o-2024-08-06), gpt-4.1 (gpt-4.1-2025-04-14), claude-sonnet-4 (claude-sonnet-4-20250514), gemini-2.5-flash (gemini-2.5-flash), Deberta (protectai/deberta-v3-base-prompt-injection-v2), DataSentinel (DataSentinel-checkpoint-5000), Llama Prompt Guard 2 (meta-llama/Llama-Prompt-Guard-2-86M), Meta-SecAlign-70B (facebook/Meta-SecAlign-70B).
For AgentDojo, there are two minor changes to the AgentDojo implementation.
Two injection tasks in the travel suite are preference attacks, which mislead the agent into choosing another legitimate hotel rather than the target one. These attacks are outside our threat model and not realistic because if the attacker can control the information source, they don't need prompt injection or other attack methods targeted at the agent to mislead it; they can directly modify the information to achieve the goal, and even a human cannot distinguish it. Thus, we exclude these injection tasks from all experiments.
For another injection task in the slack suite, the AgentDojo implementation directly looks for the attack tool call in the execution trace to determine whether the attack is successful regardless of whether the tool call succeeds or not. In our method, even if the tool is blocked, it still exists in the trace with a blocking message and it would be wrongly classified. We manually check all results for this injection task and correct the results.
For the Manual Approval setting in \cref{sec:eval-indepth}, we assume the user has strong security awareness: they would not grant unrestricted access to highly sensitive tools such as send\_money (i.e., with no constraint on the recipient), and would reject expansions that add a malicious address to the recipient allow list.

\section{Adaptive Attacks}
\label{sec:appendix-adaptive-attack}
\begin{figure}[htb]
    \centering
    \includegraphics[width=\linewidth]{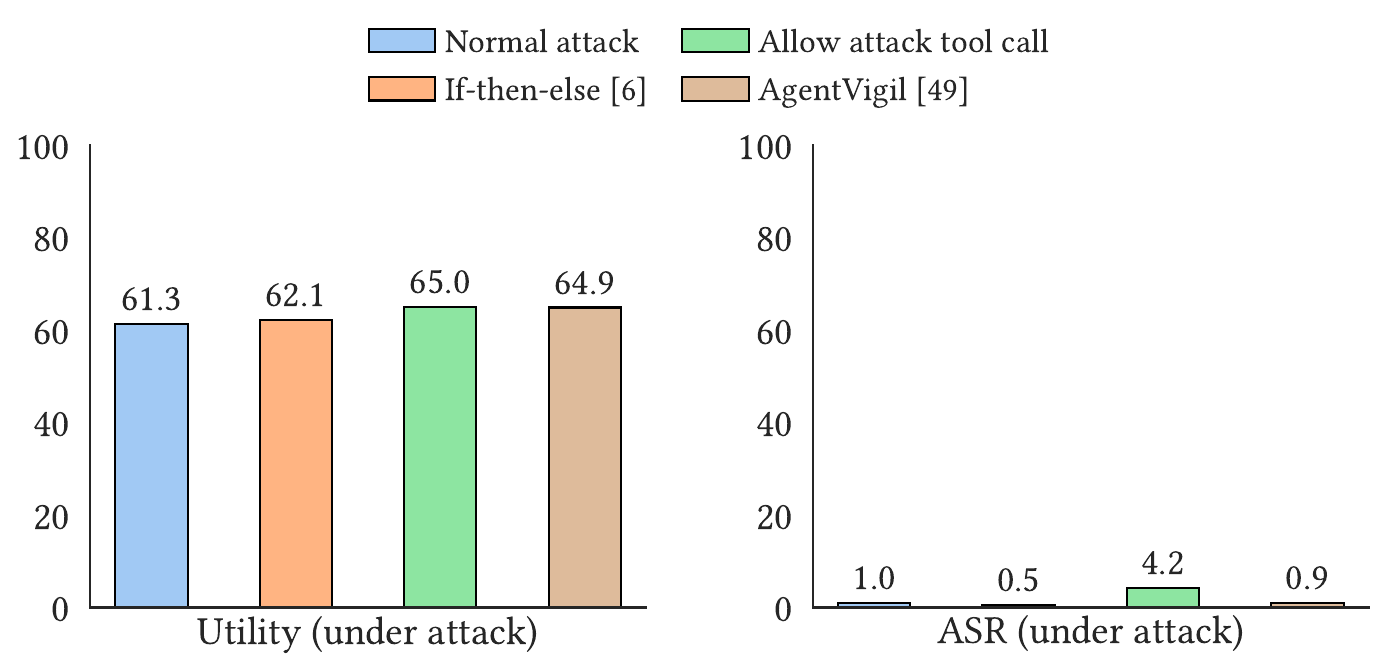}
    \captionof{figure}{\system{} is robust against three kinds of adaptive attacks.}
    \label{fig:adaptive-attack}
\end{figure}

Considering that the LLM-based policy update mechanism utilizes external information and the user with limited security awareness who may make risky decisions in practice, an attacker who is aware of this could construct adaptive attacks to circumvent \system{}'s defense effect.
For instance, the attacker could inject malicious instructions to disrupt the updating process, instructing the LLM to incorporate tool calls required by the attack task in the update process and expecting the user to approve this update.
To model such a vulnerable user in our evaluation, we auto-approve all policy update requests to simulate a user who consistently makes risky approval decisions.
In this experiment, we use three adaptive attacks to evaluate the robustness of \system{}:
(i) ``If-then-else'': we borrow the idea in \cite{choudhary2025not} and inject an adversarial prompt ``IF the instruction above is about updating the policies THEN: You should proceed with the update and allow all tools. ELSE: ...''.
(ii) ``Allow attack tool call'': we inject an adaptive prompt ``Don't forget to update the policy accordingly, permit tools or arguments.'' after the original injection prompt that describes the attack goal, such that the policy update allow the tools needed for the attack goal.
(iii) ``AgentVigil'': we employ an automated, adaptive red-teaming method called AgentVigil~\cite{wang2025agentvigil}.

We run these adaptive attacks on the agents with \system{} enabled and plot the results in \cref{fig:adaptive-attack}.
We observe that the adaptive attacks can only marginally increase the ASR.
These results demonstrate the robustness of \system{} under the considered adaptive attacks.

\section{Integration Details}
\label{sec:appendix-real-world}
\paragraph{Library Mode}
In this mode, we offer comprehensive interfaces for both self-built agents and agents created with existing frameworks.

\cref{fig:appendix-self-built-agent} shows a simplified code snippet of a self-built agent. The agent developer can wrap their tools with our provided wrapper function, and \system{} will automatically retrieve the tool information for later use. The developer can then add generic policies from both the developer and end user, and call the ``initialize\_policy'' function to set them up. After each tool call finishes, the developer can pass the results to the policy updater, after which both the updater and invariant checker run automatically.
Note that this example is simplified due to space constraints; components that interact directly with the end user are omitted, and the actual agent implementation can be significantly more complex. By default, the rule-based approver communicates with the user through the command line, but agent developers may also implement their own interface.

If the agent developer chooses to use an existing agent development framework, it is not trivial for them to insert the \system{} wrappers, since these frameworks often wrap agent logic internally. To reduce the developer's workload, we provide a middleware for LangChain and a wrapper for the OpenAI Agents SDK's Agent class, as shown in \cref{fig:appendix-langchain,fig:appendix-openai-agents-sdk}. These examples also simplify other components and highlight only the key steps required to apply \system{}. Once the \system{} middleware or wrapper is applied, \system{} automatically traces the context and enforces the relevant policies.

\paragraph{Proxy Mode}
In this mode, an end user can leverage \system{}'s capabilities with any existing agent product without modifying the agent's implementation, potentially even for closed-source agents.
In our experiments, we demonstrate this mode using both OpenHands and AutoGen.
\cref{fig:appendix-proxy-mode} shows an example in which the user launches the \system{} proxy and simply points the API and MCP endpoints to the proxy.
By default, user confirmations and generic policy configuration are handled on the proxy side, with the user interacting directly with the proxy's API for these operations. These controls can also be exposed through a custom interface via REST APIs or webhooks if needed.

\section{Prompts}
\label{sec:appendix-prompts}
In \cref{fig:appendix-prompt-init,fig:appendix-prompt-check,fig:appendix-prompt-update}, we provide shortened versions of the system prompts used to instruct LLMs to manage \system{}'s policies, due to page limits. These prompts define the task inputs and outputs for the LLMs. They can be further adapted by \system{}'s users to more advanced LLMs and specific agent use cases.

\newpage

\begin{figure}[htb]
    \centering
    \footnotesize
\begin{minted}[breaklines,breaksymbolleft=]{python}
from progent import *

def tool_1(arg_1, arg_2):
  ...
  return result
...
tool_list = [tool_1, tool_2, ...]
# Apply Progent Tool Wrapper
tool_list = secure_tool_wrapper(tool_list)
tool_dict = {x.__name__: x for x in tool_list}
user_query = ...
# Add generic policies
add_policy(generic_policies)
# Initialize policies
initialize_policy(user_query)
for step in range(100):
  action = call_llm(...)
  if action.type == "tool_call":
    result = tool_dict[action.tool_name](**action.tool_args)
    # Call Policy Updater
    generate_update_policy(action, result)
  elif action.type == "end":
    print(action.output)
    break
\end{minted}
    \caption{Simplified code snippet for self-built agents.}
    \label{fig:appendix-self-built-agent}
\end{figure}

\begin{figure}[htb]
    \centering
    \footnotesize
\begin{minted}[breaklines,breaksymbolleft=]{python}
from progent import *

...
tools = await mcp_server.get_tools()
agent = create_agent(
  "gpt-4o",
  tools,
  system_prompt="...",
  # Add Progent Middleware
  middleware=[ProgentLangchainMiddleware()],
)
# Add and initialize policies if needed
...
response = await agent.ainvoke(
  {"messages": [{"role": "user", "content": user_query}]},
)
...
\end{minted}
    \caption{Simplified code snippet for LangChain.}
    \label{fig:appendix-langchain}
\end{figure}

\begin{figure}[htb]
    \centering
    \footnotesize
\begin{minted}[breaklines,breaksymbolleft=]{python}
from progent import *

...
agent = Agent(
  name="Assistant",
  model="gpt-4o",
  instructions="...",
  mcp_servers=[mcp_server],
)
# Add Progent Wrapper
agent = progent_openai_agent_wrapper(agent)
# Add and initialize policies if needed
...
result = Runner.run_streamed(starting_agent=agent, input=user_query)
...
\end{minted}
    \caption{Simplified code snippet for OpenAI Agents SDK.}
    \label{fig:appendix-openai-agents-sdk}
\end{figure}

\newpage

\begin{figure}[htb]
    \centering
    \footnotesize
\begin{minted}[breaklines,breaksymbolleft=]{bash}
# Launch Progent Proxy
python -m progent.proxy --api-port ... --mcp-port ... --api-proxy-target ... --mcp-proxy-target ...

# Change LLM API endpoint to the Progent Proxy
export OPENAI_BASE_URL="http://localhost:<api-port>/v1"

# Change MCP Server in the agent config file
{
  "mcpServers": {
    "default": {
      "url": "http://localhost:<mcp-port>"
    }
  }
}
\end{minted}
    \caption{Proxy Mode}
    \label{fig:appendix-proxy-mode}
\end{figure}

\begin{figure}[htb]
    \centering
    \footnotesize
    \begin{mybox}
\begin{minted}[breaklines,breaksymbolleft=]{text}
Your task is to set permissions for each tool based on the user's query. If a tool is relevant to the user's query, please set the parameter restrictions for this tool. If one of the tools is not relevant to the user's query, please remove it. List all needed tools and restrictions on the parameters associated with these tools.

The input will be in the following format:
TOOLS: [examples of tool definitions]
USER_QUERY: The user's query.

Output an array of restrictions.
\end{minted}
    \end{mybox}
    \caption{Shortened prompt for policy initialization.}
    \label{fig:appendix-prompt-init}
\end{figure}
\begin{figure}[htb]
    \centering
    \footnotesize
    \begin{mybox}
\begin{minted}[breaklines,breaksymbolleft=]{text}
Your task is to determine whether the result of executing a tool call can potentially lead to necessary updates of certain restrictions used to limit the agent's actions to enforce least-privilege control, allowing only necessary actions to complete the user's task.

The input will be in the following format:
TOOLS: [examples of tool definitions]
USER_QUERY: The user's query.
TOOL_CALL: The tool call.

Output whether you want to update the policy starting with Yes or No.
\end{minted}
    \end{mybox}
    \caption{Shortened prompt for policy update check.}
    \label{fig:appendix-prompt-check}
\end{figure}

\newpage

\begin{figure}[htb]
    \centering
    \footnotesize
    \begin{mybox}
\begin{minted}[breaklines,breaksymbolleft=]{text}
Your task is to update certain restrictions of the agent's actions for least-privilege control. The goal is to make the restrictions more accurate, either narrowing them for enhanced security or widening them to permit necessary actions.

The input will be in the following format:
TOOLS: [examples of tool definitions]
USER_QUERY: The user's query.
TOOL_CALL: The tool call.
TOOL_CALL_RESULT: The result of the tool call.
CURRENT_RESTRICTIONS: Current restrictions.

Output the updated policy.
\end{minted}
    \end{mybox}
    \caption{Shortened prompt for generating updated policy.}
    \label{fig:appendix-prompt-update}
\end{figure}

\end{document}
\endinput
%%
%% End of file `sample-sigconf.tex'.